\pdfoutput=1
\documentclass[a4,11pt%,fleqn
]{article}
\usepackage{etex,geometry,epsf,amsmath,amsfonts,cite,microtype,mathtools,pdflscape,hyperref}

\hypersetup{colorlinks=true,linkcolor=blue,urlcolor=blue,citecolor=blue}

\def\ob{|} % {(}
\def\cb{|} % {)}

\def\cbI#1#2#3#4#5#6#7#8{
\setlength{\unitlength}{#1sp}%
\centering{
\begin{picture}(1500,757)(4500,-2483)
\thicklines
{\put(4850,-1860){\line( 1,0){800}}}%
{\put(4850,-2460){\line( 1,0){800}}}%
{\put(5250,-1860){\line( 0,-1){600}}}%
\put(4775,-2460){\makebox(0,0)[rc]{$#2$}}
\put(4775,-1860){\makebox(0,0)[rc]{$#3$}}
\put(5750,-1860){\makebox(0,0)[lc]{$#4$}}
\put(5750,-2460){\makebox(0,0)[lc]{$#5$}}
\put(5300,-2160){\makebox(0,0)[lc]{$#6$}}
\end{picture}}}%

\def\cbb#1#2#3#4#5#6#7#8{
\setlength{\unitlength}{#1sp}%
\begin{picture}(2800,1400)(3901,-2683)
\thicklines
{\put(4800,-1860){\line( 0,-1){600}}}%
{\put(5700,-1860){\line( 0,-1){600}}}%
{\put(4200,-2460){\line( 1, 0){2100}}}%
\put(4150,-2460){\makebox(0,0)[rc]{$#2$}}
\put(4800,-1800){\makebox(0,0)[cb]{$#3$}}
\put(4800,-2560){\makebox(0,0)[ct]{$#7$}}
\put(5250,-2380){\makebox(0,0)[cb]{$#4$}}
\put(5700,-1800){\makebox(0,0)[cb]{$#5$}}
\put(5700,-2560){\makebox(0,0)[ct]{$#8$}}
\put(6400,-2460){\makebox(0,0)[lc]{$#6$}}
\end{picture}}%

\def\fppA#1#2#3#4#5{
\setlength{\unitlength}{#1sp}%
\begin{picture}(500,450)(450,500)
\thicklines
{\put(500,500){\circle{1000}}}%
{\put(383,617){\circle*{50}}}%
{\put(617,617){\circle*{50}}}%
{\put(617,383){\circle*{50}}}%
{\put(383,383){\circle*{50}}}%
\put(330,700){\makebox(0,0)[rc]{$#2$}}
\put(700,700){\makebox(0,0)[rc]{$#3$}}
\put(700,300){\makebox(0,0)[rc]{$#4$}}
\put(330,300){\makebox(0,0)[rc]{$#5$}}
\end{picture}}

\usepackage{tikz}
\usetikzlibrary{arrows,calc,patterns,decorations.markings,decorations.pathreplacing,plotmarks,shapes.arrows,decorations.pathmorphing,backgrounds}
\tikzset{snake it/.style={decorate, decoration=snake}}

\usepackage{bbm}
\usepackage{parskip}
\usepackage{amssymb}
\usepackage{braket}
\usepackage{url}
\usepackage[all]{xy}
\usepackage{subfig}
\usepackage[normalem]{ulem}
\usepackage{rotating}
\usepackage{rotfloat}
\usepackage{array}
\numberwithin{equation}{section}
\numberwithin{table}{section}
\newcommand{\rhoo}{\kappa}

\newcommand{\vev}[1]{{\langle\;#1\;\rangle}}

\newcommand{\olT}{\overline T}
\newcommand{\nn}{\nonumber}
\newcommand{\One}{\mathbf{1}}

\newcommand{\ot}{{\ensuremath{(1,2)}}}
\newcommand{\rt}{{\ensuremath{(r,2)}}}
\newcommand{\phib}{{\bar\phi}}
\newcommand{\psib}{{\bar\psi}}
\newcommand{\vL}{{\ensuremath{\varphi_L}}}
\newcommand{\vR}{{\ensuremath{\varphi_R}}}
\newcommand{\dpp}{d}
\renewcommand{\dpp}{d_{\phi\phi}}
\newcommand{\dpbpb}{\bar d}
\renewcommand{\dpbpb}{d_{\phib\phib}}
\newcommand{\vp}{\varphi}
\newcommand{\dvp}{d_{\vp\vp}}

\def\SVIR2{{SVIR}$_3{}^{\otimes 2}$}

\newcommand{\blank}[1]{}%

\newcommand{\be}{\begin{equation}}
\newcommand{\ee}{\end{equation}}

\newcommand{\bc}{\begin{cases}}
\newcommand{\ec}{\end{cases}}

\newcommand{\cT}{{\cal T}}
\newcommand{\cR}{{\cal R}}

\newcommand{\rd}{{\mathrm{d}}}

\newcommand{\ds}{\displaystyle}

\begin{document}
\begin{flushright}  {~} \\[-12mm]
{\sf KCL-MTH-17-06 v2}
\end{flushright} 

\thispagestyle{empty}

\begin{center} \vskip 24mm
{\Large\bf 
The reflection coefficient for minimal model conformal defects from
perturbation theory.}\\[10mm]  
{\large 
Isao Makabe
and
G\'erard M.\ T.\ Watts
}
\\[5mm]
Dept.\ of Mathematics, King's College London,\\
Strand, London WC2R\;2LS, UK
\\[5mm]

\vskip 4mm
\end{center}

\vspace{2cm}
\begin{quote}{\bf Abstract}\\[1mm]
We consider a class of conformal defects in Virasoro minimal models
that have been defined as 
fixed points of the renormalisation group and calculate the leading
contribution to the reflection coefficient for these
defects. This requires several structure constants
of the operator algebra of the defect fields, for which we present a derivation in detail. We compare our results with our recent work on conformal
defects in the tricritical Ising model.
\end{quote}

{\setcounter{tocdepth}{2}
%\tableofcontents
}

\newpage

\section{Introduction}

Defects in two-dimensional systems have been studied for a long time,
see eg \cite{DMS,QRW} and references therein.
In conformal field theory, attention has been focused primarily on
defects which preserve some or all of the conformal symmetry.
If the defect lies along the real axis, this can be expressed in terms
of the continuity of various quantities.
If the holomorphic and anti-holomorphic components $T$ and $\bar T$ of
the stress-energy tensor are each separately continuous across the
defect, it is said to be topological; if $T-\bar T$ vanishes on the
defect, it is called reflecting or factorised and corresponds to some
combination of conformal boundary conditions on the upper and lower
half planes. These are both examples of the 
more general case of a conformal defect for which $T-\bar T$ is
continuous across the defect.  

The Virasoro minimal models are amongst the simplest and most
well-studied conformal field theories. The boundary conditions and
topological defects have been completely classified in 
\cite{Petkova:2000ip}
and further studied in 
\cite{FFRS}. The situation of more general conformal defects is much
less clear. The conformal defects in the Ising model were classified in  
\cite{Oshikawa:1996dj} (and in the much simpler Lee-Yang model in 
\cite{QRW}), but in general the only results found are either
perturbative or numerical \cite{krw}.
More recently, we have also found exact expressions for conformal
defects in the tricritical Ising model \cite{mw} (based on ideas in \cite{GY}).

There has also been a great deal of study of defects between different
conformal field theories, with exact classifications in a few cases
\cite{QRW}, exact proposals \cite{G1} for defects related to
renormalisation group flows, and perturbative
calculations \cite{SCB}.

One characteristic of a conformal defect is its transmission
coefficient $\cT$, or equivalently its reflection coefficient 
$\cR = 1 - \cT$, which was defined in \cite{QRW}.
These take the values $\cR=0$ for a topological defect and $\cR=1$
for a factorised defect, and $0<\cR<1$ for a general conformal defect
in a unitary theory \cite{BGLM}.

The aim of this paper is to calculate the reflection coefficient for a
class of conformal defects in Virasoro minimal models defined as the
fixed points of the perturbative renormalisation group flows considered
in \cite{krw}, and to compare this with the values found in \cite{mw}
for the tri-critical Ising model.

The structure of the paper is as follows.
In section \ref{sec2} 
we define the perturbed defects that we will consider, state their
fixed points and outline the calculation of the reflection coefficient
for these fixed points.
For this calculation we need several of the structure constants of the
operator algebra of the defect fields. These are given in \cite{RCFTIV,IR10}
in terms of topological field theory data but in section \ref{sec:scs} we
provide an alternative derivation of these constants, extending the
results of \cite{bhw}.     

In section \ref{ints} we calculate the perturbative integrals and in
section \ref{ans} we give the value of $\cR$ at the fixed points.
Finally we state our conclusions in section \ref{concs}.

%4. Set up the calculation.
%Write down the correlation functions, do the integrals, calculate
%${\cal R}$.
%
%5. Conclusions
%
%Discuss relation to known results.
%
%Discuss next-to-leading order and higher.

%\newpage

\section{The $D_\rt$ defect and its perturbations}
\label{sec2}

We will concern ourselves only with diagonal $M_{p,q}$ Virasoro
minimal models, 
also known as the $(A_{p-1},A_{q-1})$ invariant \cite{CIZ}. 
These are labelled by two co-prime integers $(p,q)$; we shall take
$p\geq 2$, $q\geq 5$.
The model has $(p-1)(q-1)/2$ primary fields corresponding to the
Virasoro highest weight representations which are
labelled by two integers $(r,s)$ with $(r,s) \simeq (p-r,q-s)$.
We are going to be especially interested in the representation
$(1,3)$, and we will write $h=h_{13}=2p/q - 1$.

The elementary topological defects for this model were classified in
\cite{Petkova:2000ip}, 
and are labelled by the same representations of the
Virasoro algebra as the
bulk fields.
The space of local fields on the defects is also known. 
If we label
the representations by $a$, then a primary field on the defect is
labelled by two representations $(a,b)$
which give its
properties under the holomorphic and anti-holomorphic copies of the
Virasoro algebra (but see the comment below on the transformation
rules for defect fields).   
The multiplicity $M_{ab}$ of
the primary field with labels $(a,b)$ on the defect with label $d$
(which is $\tilde V_{ab;d}{}^{d}$ in the notation of \cite{Petkova:2000ip})
are given in terms of the Verlinde fusion numbers $N_{abc}$ by  
\be
M_{ab}
=
\sum_e N_{dae} N_{deb}
= 
\sum_f 
N_{ddf} N_{fab}
\;.
\label{mult}
\ee

From the formula \eqref{mult}, a general $(r,s)$ defect has (for $s>2$
and $q$ large enough) one chiral 
field of weights $(h,0)$, one field of weights $(0,h)$ and three 
%>>>>>>>>>>>>>>>>>>
%<<<<<<<<<<<<<<<<<<
fields of weight $(h,h)$. 
A defect of type $(r,2)$ is special in that it has one chiral
field $\phi$ of conformal weights $(h,0)$, one chiral field $\phib$ of
weights $(0,h)$, but only a two dimensional space of fields
$\{\vp_\alpha\}$ of weights $(h,h)$.  

Furthermore, the $\rt$ topological defect can be constructed as the
fusion $(r,1)$ and $(1,2)$ topological defects and the operator
product algebra of fields of type $(a,b)=((1,s)(1,s'))$ is unaffected
by this fusion, in exactly the same way that the action of
topological defects on boundaries leaves operator algebras invariant
\cite{Graham}. This means that when considering the algebra of fields generated by the set $\{\One,\phi,\bar\phi,\varphi_\alpha\}$, 
we can restrict attention to just the
$(1,2)$ defect.

The fact that there is a two-dimensional space of fields
$\{\vp_a\}$ on the $(r,2)$ defects allows one to choose 
a canonical basis of these fields with special
properties so that the analysis of the sewing constraints is
correspondingly simpler.
These sewing constraints have been solved in \cite{bhw}
for the $\ot$ defect in the non-unitary Lee-Yang model, the
$(A_1,A_4)$ theory, in which $D_\ot$ is the only non-trivial defect
and $\{\One,\phi,\phib,\vp_\alpha\}$ are the only non-trivial primary
defect fields. In this paper we extend this analysis to the 
fields $\{\One,\phi,\phib,\vp_\alpha\}$ on defects of type $D_\rt$ in
all the $(A_p,A_q)$ models.

We are interested in the perturbations of the defect $D_\rt$ by a
combination of the fields $\phi$ and $\phib$, 
\be
S =  \int \,\left( \lambda \phi(x) + \bar\lambda \phib(x) \right) \,
\rd x
\;.
\label{eq:p}
\ee
where
the parameters $\lambda$ and $\bar\lambda$ are independent.
This is a relevant perturbation if $h<1$ which is the case if $p<q$.

One important question is
that of the transformation properties of fields on a defect under a conformal
transformation. 
%One can define a standard position for the defect to be along the real axis and
%in any other position we can say they are given by the result of a conformal map
%from the defect in standard position to the new position. 
%The question is: how do defect fields transform under such a map?
%The defect defines a direction (the
%tangent vector along the defect) through the insertion point of the field,
%and so a defect field can pick up an extra phase under a conformal
%transformation. 
We will use the conventions of \cite{IR10} which imply that defect
fields always transform with the absolute value of the derivative of
the conformal map, even if they are ``chiral'' defect fields. This is
possible because the defect defines a direction  through the insertion
point of the field (the tangent vector along the defect), and so a
defect field can pick up an extra phase under a conformal
transformation: this is 
chosen so that all defect fields transform with the absolute value of
the derivative of the conformal map. This has the advantage of making
the perturbation well-defined on defects that are closed loops and
making the correlation function independent of the orientation
of the defect at the location of the defect field (as one would expect
if the defect is genuinely topological).  
The question remains whether this choice for the transformation law of
``chiral'' defect fields is unique: the corresponding situation for a
boundary and boundary fields was considered by Runkel  
\cite{runkelPhD}, and there seems no way to fix it %this phase 
a priori; we stick to the conventions of  \cite{IR10} here for the
good reasons cited above.
%>>>>>>>>>>>>>>>
%<<<<<<<<<<<<<<<

The expectation  
values in the perturbed defect $D_\rt(\lambda,\bar\lambda)$ are
formally given by
\be
 \langle\, {\cal O} \,\rangle_{D_\rt(\lambda,\bar\lambda)}
= 
 \langle\, {\cal O} \,
 \exp( - S )   \,\rangle_{D_\rt}
\;.
\label{eq:dp}
\ee
This is only formal since there may be UV divergences in the integrals when
the insertion points of two fields $\phi$ or two fields $\phib$
meet and IR divergences from integration along the whole real axis.
This means that the general procedure of regularisation and
renormalisation may be needed to given meaning to the expression
\eqref{eq:dp}. This is explained in Affleck and Ludwig
\cite{Affleck:1991tk} and applied by Recknagel et al in
\cite{Recknagel:2000ri} to the case of boundary perturbations of the
unitary minimal models where $q=p+1$.

As explained in \cite{krw}, when $y=1-h$ is small and positive, the results of
\cite{Recknagel:2000ri} can immediately be applied to the case of
defects with the perturbation \eqref{eq:p} with the prediction (from
third order perturbation theory) of three conformal defects at the
fixed points  
\begin{align}
\mathrm{ (i)}\;\; & \lambda = \lambda^*, \bar \lambda = 0
\\
\mathrm{(ii)}\;\; & \lambda = 0, \bar \lambda = \lambda^*
\\
\mathrm{(iii)}\;\; & \lambda = \bar \lambda = \lambda^*
\end{align}
The fixed points (i) and (ii) can be identified as the defect
$D_{(2,1)}$ (if $r=2$) and (more generally) the superposition
$D_{(r-1,1)}\oplus D_{(r+1,1)}$; the fixed point (iii) is a 
potential new conformal defect, denoted by $C$ in \cite{krw} in the
case of the perturbation of the defect $D_\ot$.
The value of $\lambda^*$ is given (to first order in $y$) by
\be
 \lambda^* 
= \frac{y}{C_{\phi\phi\phi}}
= \frac{y}{C_{\phi\phi}^\phi d_{\phi\phi}}
\;,
\ee
where $C_{\phi\phi}^\phi$ is the three point coupling of the fields $\phi$.
Note that $\lambda^*$ depends on the normalisation of $\phi$, but this
will cancel in any physical quantities.

%\newpage

\subsection{The perturbative calculation of the reflection and transmission coefficients}

The transmission and reflection coefficients of a conformal defect
along the real axis
were defined in \cite{QRW} as 
\be
 \cR
= \frac{\langle T^1 \olT{}^1 + T^2 \olT{}^2\rangle}
       {\langle (T^1 + \olT{}^2)( \olT{}^1 + T^2)\rangle}
\;,\;\;
  \cT = 1 - \cR
\;
\ee
where $T^1$ and $\olT{}^1$ are inserted at the point $iY$ on the upper
half-plane, while $T ^2$ and $\olT{}^2$ are inserted at the point $-iY$.
For the unperturbed topological defect, 
\be
  \vev{ T^1 \olT{}^1}
= \vev{ T^2 \olT{}^2} =0 
\;,\;\;
  \vev{ T^1 T^2}
= \vev{ \olT{}^1 \olT{}^2} = \frac{c}{32 Y^4}
\;,
\ee
and so  
$\cR=0$ and $\cT=1$.

For the defect with perturbation \eqref{eq:p}, the expansion of the
perturbed quantities using
\eqref{eq:dp} gives
\begin{align}
  \vev{ T^1 \olT{}^1}
&= 
\frac{1}{4}
  \lambda^2\bar \lambda^2\int \rd x\, \rd x'\, \rd y\, \rd y'
  \vev{  T(iY)\,\olT(iY)\,  \phi(x) \phi(x') \bar\phi(y)  \bar\phi(y') }
\nonumber\\
&-
\frac{1}{24}
  \lambda^3\bar \lambda^2\int \rd x\, \rd x'\, \rd x''\, \rd y\, \rd y'
  \vev{T(iY)\,\olT(iY)\,  \phi(x) \phi(x') \phi(x'') \bar\phi(y)  \bar\phi(y')}
\nonumber\\
&-
\frac{1}{24}
  \lambda^2\bar \lambda^3\int \rd x\, \rd x'\, \rd y\, \rd y'\, \rd y''
  \vev{T(iY)\,\olT(iY)\,\phi(x) \phi(x')\bar\phi(y) \bar\phi(y') \bar\phi(y'')}
\nonumber\\
&+
O(\lambda^6)
\;,\;\;
\label{eq:T1Tb1}\\
  \vev{ T^1 T^2}
&= \frac{c}{32 Y^4}
\nonumber\\
&+
  \frac{1}{2}\lambda^2\int \rd x \,\rd x' 
  \vev{T(iY)\,T(-iY)\,\phi(x) \phi(x') }
\nonumber\\
&+
  \frac{1}{2}\bar \lambda^2\int \rd y\, \rd y'
  \vev{T(iY)\,T(-iY)\,\bar\phi(y)  \bar\phi(y') }
 + O(\lambda^3)
\;,
\end{align}
and so to find the leading order term in $\cR$, we only need to calculate
the first term in $\vev{ T^1 \olT{}^1}$
and  $\vev{ T^2 \olT{}^2}$.
It turns out there are neither UV nor IR divergences in these
integrals, their dependence on $Y$ is simply $Y^{-4}$ and
the reflection coefficient $\cR$ (to leading order) is indeed
independent of $Y$ as expected. We shall take $Y=1$ from now on.

The consequence is that the only correlation function we need to
evaluate is
\be
  \vev{ T(i)\olT(i) \phi(x) \phi(x') \bar\phi(y)  \bar\phi(y') }
\;,
\ee
where the insertion points can be in any order.
This is equal to 
\be
  \vev{ T(-i)\olT(-i) \phi(-x) \phi(-x') \bar\phi(-y)  \bar\phi(-y') }
\;,
\ee
by rotation through $\pi$.

The analytic structure is simple, 
\be
\left\langle
\;
T(i)\, \olT(i) 
\, \phi(x) \phi(x') \phib(y) \phib(y')
\;
\right\rangle
= \mathcal{C}\, \frac{(x'-x)^{2-2h}\,(y'-y)^{2-2h}}
               {(i-x)^2(i-x')^2(i+y)^2(i+y')^2 } 
\;,
\ee
but the constant $\mathcal{C}$ depends on the order of the insertion
points $\{x,x',y,y'\}$ and is determined by the operator algebra
structure constants, so we now turn to the calculation of some of the
structure constants of the local fields on the defect $D_{\rt}$.

%\newpage

\section{The structure constants}
\label{sec:scs}

In this section we will calculate some structure constants for the
$(r,2)$ defect in the diagonal Virasoro Minimal models. 
These structure constants can be found in terms of
topological field theory data \cite{RCFTIV,IR10} which is a  general
method allowing one to find all the structure constants in
the defect theory, but we will not use it here and instead only use
elementary properties of the conformal field theory to find the
particular structure constants we need for the perturbative 
calculation of the reflection coefficient in the minimal
models. 

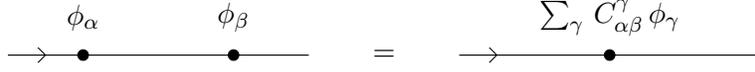
\begin{figure}[htb]
\[
  \begin{tikzpicture}

\draw (-5,0) -- (-1,0);
\draw (1,0) -- (5,0);

\draw (1.4,.1) -- (1.5,0) -- (1.4,-.1);
\draw (-4.6,.1) -- (-4.5,0) -- (-4.6,-.1);

\filldraw (-4,0) circle (2pt);
\filldraw (-2,0) circle (2pt);
\filldraw (3,0) circle (2pt);

\node at (-4,0.5) {$\phi_\alpha$} ;
\node at (-2,0.5) {$\phi_\beta$} ;

\node at (0,0) {$=$} ;

\node at (3,0.5) {$\sum_\gamma\,C_{\alpha\beta}^\gamma\,\phi_\gamma$} ;

  \end{tikzpicture}
\]
\caption{The OPE of defect fields}
\label{fig:OPE}
\end{figure}

We note here that we will use the conventions of \cite{IR10} so that
the structure constant $C_{\alpha\beta}^\gamma$ is the coefficient of
the field $\phi_\gamma$ appearing in the OPE of the fields
$\phi_\alpha(x)$ with $\phi_\beta(y)$ on the defect oriented opposite
to the real line with $x>y$, which
% Rotating by $\pi$ this 
means that this
coefficient appears in the OPE of the fields $\phi_\alpha$ with
$\phi_\beta$ as they appear along the defect. Rotating by $\pi$, we
obtain the picture in figure \ref{fig:OPE}.

%
%We will calculate the structure constants by solving the sewing
%constraints for the defect fields. The reason we consider only the
%$(r,2)$ defects is that there is a distinguished basis of fields on
%the defect which makes the analysis of the sewing constraints
%substantially easier. 
%

\subsection{The bulk theory}

The  $(A_{p-1},A_{q-1})$ Virasoro minimal model has $(p-1)(q-1)/2$
bulk primary fields, of which we are especially interested in the field
$\vp$ of type $(1,3)$.
If we set $t=p/q$, then $h_{1,3} = h = 2t-1$ and $h<1$ if $t<1$, that
is $p<q$.

The fusion rules for this field are
\be
 [\vp] \star [\vp] = [1] + [\vp] + [\chi]
\;,
\ee
where $\chi$ is of type (1,5) and has conformal weights $(h',h')$
where $h' = h_{1,5} = 6t-2$. Hence, the OPE of $\vp$ with itself is 
\be
\varphi(z,\bar z) \varphi(w,\bar w) =
  \frac{d_{\varphi\varphi}}{|z-w|^{4h}}
+ \frac{C_{\varphi\varphi}^\varphi \, \varphi(w,\bar w)}{|z-w|^{2h}}
+ \frac{C_{\varphi\varphi}^\chi \, \chi(w,\bar w)}{|z-w|^{4h-2h'}}
+ \ldots
\label{eq:bope}
\ee
The structure constant $C_{\vp\vp}^\vp$ clearly depends on the choice
of $d_{\vp\vp}$ (see eg \cite{DF,runkel} for different conventions)
but the combination 
\be
 \frac{(C_{\vp\vp}^{\vp})^2}{{d_{\vp\vp}}}
= -(1 - 2t)^2
  \frac{\Gamma(2 - 3t)}{\Gamma(3t-1)}
  \frac{\Gamma(4t-1)^2}{\Gamma(2 - 4t)^2}
  \frac{\Gamma(t)^3}{\Gamma(1 - t)^3}
  \frac{\Gamma(1-2t)^4}{\Gamma(2t)^4}
\;,
\label{eq:Cppp}
\ee
is independent of the normalisation.

If $h=1-y$ then 
\be
 \frac{(C_{\vp\vp}^{\vp})^2}{{d_{\vp\vp}}}
= \frac{16}3 - 16 y + O(y)^2
\;.
\label{eq:Cppp2}
\ee

\subsection{The defect theory}

The defects of the  $(A_{p-1},A_{q-1})$ Virasoro models are not
intrinsically oriented, but the operator product of fields along
the defect depends on the ordering of the fields, we shall assume that
we can define an orientation for the defects but that all results will
be independent of this orientation.

Since the space of fields $\{\varphi_\alpha\}$ of weights $(h,h)$ is
only two-dimensional for a defect of type \rt, we can take as a basis the
fields $\vL$ and $\vR$ which are the limits of the bulk field $\varphi$ as it
approaches the defect from the left or the right respectively as one
looks along 
the defects - see figure \ref{fig:LR}.

\begin{figure}[htb]
\[
  \begin{tikzpicture}

\draw (-5,0) -- (-1,0);
\draw (1,0) -- (5,0);

\draw (1.9,.1) -- (2,0) -- (1.9,-.1);
\draw (-4.1,.1) -- (-4,0) -- (-4.1,-.1);

\draw[->] (-3,1) -- (-3,0.3);
\draw[->] (3,-1) -- (3,-0.3);

\filldraw (-3,0) circle (2pt);
\filldraw (3,0) circle (2pt);

\node at (-3,1.3) {$\vp(x+iy)$} ;
\node at (-3,-.3) {$\vp_L(x)$} ;

\node at (3,-1.3) {$\vp(x-iy)$} ;
\node at (3,.3) {$\vp_R(x)$} ;

  \end{tikzpicture}
\]
\caption{The fields $\vL$ and $\vR$ defined as limits of the bulk field}
\label{fig:LR}
\end{figure}
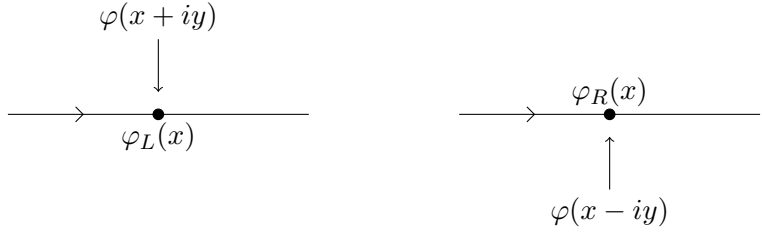

\blank{
\begin{figure}[htb]
\[
  \begin{picture}(450,140)
    \put(0,0){\scalebox{.8}{\includegraphics{defectfields_cropped.eps}}}
  \end{picture}
\]
\caption{The fields $\vL$ and $\vR$ defined as limits of the bulk field}
\label{fig:LR}
\end{figure}
}

Note that the operator product algebra of the fields
$\{\One,\phi,\phib,\vp_L,\vp_R\}$ does not close on these fields,
other fields can arise as well, namely
fields with weights $(h,h')$, $(h',h)$ and $(h',h')$ which we denote
by $\psi$, $\psib$ and $\{\chi_{L},\chi_R\}$ (which again are the
limits of the field $\chi(z,\bar z)$ as it approaches the defect from
the left and the right).
Although we should mention the existence of these fields and their
occurrence in the operator products of some of the fields
$\{\phi,\phib,\vp_\alpha\}$, we will not need any of the structure
constants including these fields as they will not contribute to any of
the sewing constraints considered later on.

We use the generic labels $\{a,b,..\}$ for all of these fields and the
labels $\{\alpha,\beta,..\}$ for the set $\{L,R\}$. The conformal
weights of the field $\Phi_a$ are $(h_a,\bar h_a)$ as in table
\ref{tab1}

\begin{table}[htb]
\be
\begin{array}{r|cccccc}
\Phi_a & h_a & \bar h_a \\
\hline
\One & 0 & 0 \\
\phi & h & 0 \\
\phib & 0 & h \\
\vp_\alpha & h & h \\
\psi & h' & h \\
\psib & h & h' \\
\chi_\alpha & h' & h' \\
\hline
\end{array}
\ee
\caption{Some of the primary fields occurring on the defect $(r,2)$}
\label{tab1}
\end{table}

We now define the structure constants between these fields from their
operator product expansions (we show the possibility of fields
$\{\psi,\psib,\chi_\alpha\}$ appearing in an OPE by placing the fields
in square brackets $[~~]$).

If both fields chiral, there are 8 structure constants
$\{\dpp,\dpbpb,C_{\phi\phi}^\phi,C_{\phib\phib}^\phib,C_{\phi\phib}^\alpha,C_{\phib\phi}^\alpha\}$
appearing in the OPEs (recall here that $x$ and $y$ are ordered along
the defect) : 
\begin{align}
\phi(x) \phi(y) &= 
  \frac{\dpp}{\ob x-y \cb^{2h}} 
+ \frac{  C_{\phi\phi}^\phi\,\phi(y) }{\ob x-y \cb^h}
 + \ldots \\
\phib(x) \phib(y) &= 
  \frac{\dpbpb}{\ob x-y \cb^{2h}} 
+ \frac{  C_{\phib\phib}^\phib\,\phib(y) }{\ob x-y \cb^h}
 + \ldots 
\;,
\\
\phi(x) \phib(y) &= 
  C_{\phi\phib}^L\,\vL(x,y)
+
  C_{\phi\phib}^R\,\vR(x,y)
 + \ldots 
\;,
\\
\phib(x) \phi(y) &= 
  C_{\phib\phi}^L\,\vL(y,x)
+
  C_{\phib\phi}^R\,\vR(y,x)
 + \ldots 
\;.
\end{align}
%where
%\[
%\dpp = d_{\phi\phi}
%\;,\;\;
%\dpbpb = d_{\phib\phib}
%\;,\;\;
%\]

With one chiral field on the left, there are 12 structure constants
$\{
C_{\phi\alpha}^\phib,
C_{\phib\alpha}^\phi,
C_{\phi\alpha}^\beta,
C_{\phib\alpha}^\beta
\}$
in the OPEs
\begin{align}
\phi(x) \varphi_\alpha(z,\bar z) &= 
  \frac{  C_{\phi \alpha}^\phib\, \phib(\bar z) }{\ob x-z \cb^{2h}}
+ \frac{  C_{\phi \alpha}^L   \, \vL(z,\bar z) }{\ob x-z \cb^{h}}
+ \frac{  C_{\phi \alpha}^R   \, \vR(z,\bar z) }{\ob x-z \cb^{h}}
+ [\psi] + \ldots 
\;,
\\
%
%\phi(x) \vR(z,\bar z) &= 
%  \frac{  C_{\phi R}^\phib\, \phib(\bar z) }{(x-z)^{2h}}
%+ \frac{  C_{\phi R}^L   \, \vL(z,\bar z) }{(x-z)^{h}}
%+ \frac{  C_{\phi R}^R   \, \vR(z,\bar z) }{(x-z)^{h}}
%+ [\psi] + \ldots \\
%
\phib(x) \varphi_\alpha(z,\bar z) &= 
  \frac{  C_{\phib \alpha}^\phi\, \phi(z) }{\ob x-\bar z \cb^{2h}}
+ \frac{  C_{\phib \alpha}^L  \, \vL(z,\bar z) }{\ob x-\bar z \cb^{h}}
+ \frac{  C_{\phib \alpha}^R  \, \vR(z,\bar z) }{\ob x-\bar z \cb^{h}}
+ [\psib] + \ldots 
\;.
%\\
%
%\phib(x) \vR(z,\bar z) &= 
%  \frac{  C_{\phib R}^\phi\, \phi( z) }{(x-\bar z)^{2h}}
%+ \frac{  C_{\phib R}^L  \, \vL(z,\bar z) }{(x-\bar z)^{h}}
%+ \frac{  C_{\phib R}^R  \, \vR(z,\bar z) }{(x-\bar z)^{h}}
%+ [\psib] + \ldots 
%
\end{align}

likewise there are 12 structure constants
$\{
C_{\alpha\phi}^\phib,
C_{\alpha\phib}^\phi,
C_{\alpha\phi}^\beta,
C_{\alpha\phib}^\beta
\}$
in the OPEs with 
one field chiral on the right:
\begin{align}
\varphi_\alpha(z,\bar z) \phi(x)  &= 
  \frac{  C_{\alpha \phi}^\phib\, \phib(\bar z) }{\ob z-x \cb^{2h}}
+ \frac{  C_{\alpha \phi}^L   \, \vL(z,\bar z) }{\ob z-x \cb^{h}}
+ \frac{  C_{\alpha \phi}^R   \, \vR(z,\bar z) }{\ob z-x \cb^{h}}
+ [\psi] + \ldots 
\;,
\\
%
%\vR(z,\bar z) \phi(x)  &= 
%  \frac{  C_{R \phi}^\phib\, \phib(\bar z) }{(z-x)^{2h}}
%+ \frac{  C_{R \phi}^L   \, \vL(z,\bar z) }{(z-x)^{h}}
%+ \frac{  C_{R \phi}^R   \, \vR(z,\bar z) }{(z-x)^{h}}
%+ [\psi] + \ldots \\
%
\varphi_\alpha(z,\bar z) \phib(x)  &= 
  \frac{  C_{\alpha \phib}^\phi\, \phi( z) }{\ob \bar z-x \cb^{2h}}
+ \frac{  C_{\alpha \phib}^L   \, \vL(z,\bar z) }{\ob \bar z-x \cb^{h}}
+ \frac{  C_{\alpha \phib}^R   \, \vR(z,\bar z) }{\ob \bar z-x \cb^{h}}
+ [\psib] + \ldots 
\;.
%\\
%
%\vR(z,\bar z) \phib(x)  &= 
%  \frac{  C_{R \phib}^\phi\, \phi( z) }{(\bar z-x)^{2h}}
%+ \frac{  C_{R \phib}^L   \, \vL(z,\bar z) }{(\bar z-x)^{h}}
%+ \frac{  C_{R \phib}^R   \, \vR(z,\bar z) }{(\bar z-x)^{h}}
%+ [\psib] + \ldots 
%
\end{align}
Finally there are 20 structure constants
$\{
d_{\alpha\beta},
C_{\alpha\beta}^\phi,
C_{\alpha\beta}^\phib,
C_{\alpha\beta}^\gamma
\}$ in the OPEs involving no chiral fields:
\begin{align}
\varphi_\alpha(z,\bar z) \varphi_\beta(w,\bar w) &= 
  \frac{d_{\alpha\beta}}{|z-w|^{4h}}
+ \frac{C_{\alpha\beta}^\phi \phi(w)}{\ob z-w \cb^h \ob\bar z - \bar w \cb^{2h}}
+ \frac{C_{\alpha\beta}^\phib \phib(\bar w)}{\ob \bar z-\bar w \cb^h \ob z - w \cb^{2h}}
\nn\\&
+ \frac{C_{\alpha\beta}^L \vL(w,\bar w)}{|z-w|^{2h}}
+ \frac{C_{\alpha\beta}^R \vR(w,\bar w)}{|z-w|^{2h}}
+ [\psi,\psib,\chi_\alpha] + \ldots
\;.
%\\
%\vL(z,\bar z) \vR(w,\bar w) &= 
%  \frac{d_{LR}}{|z-w|^{4h}}
%+ \frac{C_{LR}^\phi \phi(w)}{(z-w)^h (\bar z - \bar w)^{2h}}
%+ \frac{C_{LR}^\phib \phib(\bar w)}{(\bar z-\bar w)^h ( z - w)^{2h}}
%\nn\\&
%+ \frac{C_{LR}^L \vL(w,\bar w)}{|z-w|^{2h}}
%+ \frac{C_{LR}^R \vR(w,\bar w)}{|z-w|^{2h}}
%+ [\psi,\psib,\chi_\alpha] + \ldots
%\\
%\vR(z,\bar z) \vL(w,\bar w) &= 
%  \frac{d_{RL}}{|z-w|^{4h}}
%+ \frac{C_{RL}^\phi \phi(w)}{(z-w)^h (\bar z - \bar w)^{2h}}
%+ \frac{C_{RL}^\phib \phib(\bar w)}{(\bar z-\bar w)^h ( z - w)^{2h}}
%\nn\\&
%+ \frac{C_{RL}^L \vL(w,\bar w)}{|z-w|^{2h}}
%+ \frac{C_{RL}^R \vR(w,\bar w)}{|z-w|^{2h}}
%+ [\psi,\psib,\chi_\alpha] + \ldots
%\\
%\vR(z,\bar z) \vR(w,\bar w) &= 
%  \frac{d_{RR}}{|z-w|^{4h}}
%+ \frac{C_{RR}^\phi \phi(w)}{(z-w)^h (\bar z - \bar w)^{2h}}
%+ \frac{C_{RR}^\phib \phib(\bar w)}{(\bar z-\bar w)^h ( z - w)^{2h}}
%\nn\\&
%+ \frac{C_{RR}^L \vL(w,\bar w)}{|z-w|^{2h}}
%+ \frac{C_{RR}^R \vR(w,\bar w)}{|z-w|^{2h}}
%+ [\psi,\psib,\chi_\alpha] + \ldots
\end{align}

Having defined the fifty-two structure constants we need to calculate,
we now set about finding relations. The simplest come from the fact
that the orientation of the defect is in fact not physical.

\subsection{Symmetry relations}

Since the defect is not intrinsically oriented, our labelling
over-counts the structure constants: 
sixteen constants are related by
changing the orientation of the defect, as follows:
\be
  C_{\phi\phib}^L = C_{\phib\phi}^R
\;,\;\;
  C_{\phi\phib}^R = C_{\phib\phi}^L
\;,\;\;
 d_{LL} = d_{RR}
\;,\;\;
 d_{LR} = d_{RL}
 \;,\;\;
\ee\be
 C_{LL}^L = C_{RR}^R
\;,\;\;
 C_{LL}^R = C_{RR}^L
\;,\;\;
 C_{LR}^L = C_{RL}^R
\;,\;\;
 C_{RL}^L = C_{LR}^R
\;.
\ee

\be
 C_{\phi R}^R = C_{L\phi}^L
\;,\;\;
 C_{\phi R}^L = C_{L\phi}^R
\;,\;\;
 C_{\phi L}^R = C_{R\phi}^L
\;,\;\;
 C_{\phi L}^L = C_{R\phi}^R
\;,\;\;
\ee\be
 C_{\phib R}^R = C_{L\phib}^L
\;,\;\;
 C_{\phib R}^L = C_{L\phib}^R
\;,\;\;
 C_{\phib L}^R = C_{R\phib}^L
\;,\;\;
 C_{\phib L}^L = C_{R\phib}^R
\;.\;\;
\ee

\subsubsection{Bulk field relations}

We can use the fact that \vL\ and \vR\ are the limits of bulk fields to
find $d_{LL}$, $d_{LR}$, $d_{RL}$ and $d_{RR}$, as well as $C_{LL}^L$,
$C_{LL}^R$, $C_{RR}^L$ and $C_{RR}^R$.

In the bulk, we have \eqref{eq:bope}.
Bringing this OPE towards a defect from the left, we obtain
\be
 d_{LL} = d_{\varphi\varphi}
\;,\;\;
 C_{LL}^L = C_{\varphi\varphi}^\varphi
\;,\;\;
 C_{LL}^R = C_{LL}^\phi = C_{LL}^\phib = 0
\;,
\label{eq:319}
\ee
We have also found that 
\be
 C_{LL}^{\chi_L^{\phantom{\phi}}} = C_{\vp\vp}^\chi
\;,\;\;
 C_{LL}^{\chi_R^{\phantom{\phi}}} = C_{LL}^\psi = C_{LL}^\psib = 0
\;,
\ee
but these four constants are not of interest to us.

Likewise, bringing the bulk OPE \eqref{eq:bope} towards a defect from
the right, we obtain  
\be
 d_{RR} = d_{\varphi\varphi}
\;,\;\;
 C_{RR}^R = C_{\varphi\varphi}^\varphi
\;,\;\;
 C_{RR}^L  = C_{RR}^\phi = C_{RR}^\phib = 0
\;.
\label{eq:321}
\ee

Finally, using the expression for the defect in terms of projectors
\cite{Petkova:2000ip}
\be
 \hat D_{r,2} = \sum_{r',s} \frac{S_{(r,2),(r',s)}}{S_{(1,1),(r',s)}}
 \hat P_{r',s}
\;,
\ee
where $S_{(rs)(r's')}$ is the modular S-matrix given in the appendix,
we have
\begin{align}
 d_{LR} &= 
  \frac{ \bra{\varphi} \, \hat D_{r,2} \, \ket{\varphi}}
       {\bra 0 \,  \hat D_{r,2}\, \ket 0} 
= \frac{  S_{(r,2),(1,3)} / S_{(1,1),(1,3)} }
        {  S_{(r,2),(1,1)} / S_{(1,1),(1,1)} }
\,
 \frac{\langle{\varphi}\vert{\varphi}\rangle}
       {\langle{ 0 }\vert{ 0 }\rangle}
  \nn\\
&= ( 2 \cos(2 \pi t) -1 ) \, d_{LL}
\nn\\
&= \gamma \; d_{LL}
\;,
\label{eq:dlr}
\end{align}
where we define
\be
\gamma = 2 \cos(2 \pi t) -1
\;.
\ee
which is independent of $r$, as expected.

\subsection{Defect -- boundary identification}

We next use the fact that the OPE algebra of $\phi$ along the real
axis is the same as that of the boundary field on the \rt\ boundary -
we obtain this identification by bringing the \rt\ defect next to the
identity boundary as considered in \cite{Graham}. Likewise, the
algebra of $\phib$ is also the same as the boundary algebra. 

This means that
\be
  d_{\phi\phi} = d_{\phib\phib} 
\;,\;\;
 C_{\phi\phi}^\phi = C_{\phib\phib}^\phib
\;,
\label{bdid}
\ee
and these values are are given by Runkel's solution to the boundary
algebra \cite{runkel},
\be
\frac{(C_{\phi\phi}^\phi)^2}{d_{\phi\phi}}
=
 \frac{\Gamma(2 - 3t) \Gamma(t) \Gamma(1 - 2t)^3}
      {\Gamma(2 - 4t)^2 \Gamma(-1 + 2t) \Gamma(1 -t)^2}
\;.
\ee
If $h=1-y$ then
\be
\frac{(C_{\phi\phi}^\phi)^2}{d_{\phi\phi}}
= \frac{8}{3} - 4 y + O(y^2)
\;.
\label{eq:327}
\ee
Note that the structure constant again does not depend on $r$.

\subsubsection{Three-point function constraints}

We can express the three point function
\be
\langle \Phi_a(u) \Phi_b(v) \Phi_c(w) \rangle
\;,
\ee
in two different ways, using the OPE of $\Phi_a$ with $\Phi_b$ first,
or instead using the OPE of $\Phi_b$ with $\Phi_c$ first, 
leading to the constraint
\be
\sum_e  C_{ab}^e d_{ec} = \sum_f d_{af} C_{bc}^f
\;.
\ee

Taking $a$ and $c$ chiral, this gives the simple relations
\be
C_{\phi R}^\phib d_{\phib\phib} = C_{R\phib}^\phi d_{\phi\phi}
\;,\;\;
C_{\phi L}^\phib d_{\phib\phib} = C_{L\phib}^\phi d_{\phi\phi}
\;,\;\;
\ee\be
C_{\phib R}^\phi d_{\phi\phi} = C_{R\phi}^\phib d_{\phib\phib}
\;,\;\;
C_{\phib L}^\phi d_{\phi\phi} = C_{L\phi}^\phib d_{\phib\phib}
\;,\;\;
\ee
which, using \eqref{bdid} become
\be
C_{\phi R}^\phib = C_{R\phib}^\phi 
\;,\;\;
C_{\phi L}^\phib = C_{L\phib}^\phi
\;,\;\;
C_{\phib R}^\phi = C_{R\phi}^\phib 
\;,\;\;
C_{\phib L}^\phi = C_{L\phi}^\phib 
\;.\;\;
\ee

Taking only $a$ chiral and the two non-chiral fields equal, this gives
the slightly more complicated 
\be
 C_{\phi R}^R d_{RR} + C_{\phi R}^L d_{LR} = C_{RR}^\phi d_{\phi\phi} = 0
\;,
 C_{\phib R}^R d_{RR} + C_{\phib R}^L d_{LR} = C_{RR}^\phib
 d_{\phib\phib} = 0
\;,\;\;
\ee\be
 C_{\phi L}^R d_{RL} + C_{\phi L}^L d_{LL} = C_{LL}^\phi d_{\phi\phi} = 0
\;,
 C_{\phib L}^R d_{RL} + C_{\phib L}^L d_{LL} = C_{LL}^\phib d_{\phib\phib} = 0
\;,
\ee
which using \eqref{eq:dlr}
become
\be
 C_{\phi R}^R = - \gamma C_{\phi R}^L 
\;,
 C_{\phib R}^R = - \gamma C_{\phib R}^L
\;,\;\;
C_{\phi L}^L = - \gamma C_{\phi L}^R  
\;,
C_{\phib L}^L =  - \gamma C_{\phib L}^R 
\;.
\ee

Taking $a$ chiral and the other two fields different, we get
\be
 C_{LR}^\phi d_{\phi\phi} 
= d_{LL} C_{R\phi}^L + d_{LR} C_{R\phi}^R
\;,
 C_{LR}^\phib d_{\phib\phib} 
= d_{LL} C_{R\phib}^L + d_{LR} C_{R\phib}^R
\;,
\ee\be
 C_{RL}^\phi d_{\phi\phi} 
= d_{RR} C_{R\phi}^R + d_{RL} C_{R\phi}^L
\;,
 C_{RL}^\phib d_{\phib\phib} 
= d_{RR} C_{R\phib}^R + d_{RL} C_{R\phib}^L
\;.
\ee
Using $d_{LR} = \gamma \dvp$, these become
\be
  C_{LR}^\phi 
= \frac{\dvp}{d_{\phi\phi}}
  (C_{R\phi}^L + \gamma C_{R\phi}^R)
\;,\;\;
  C_{LR}^\phib 
= \frac{\dvp}{d_{\phi\phi}}
  (C_{R\phib}^L + \gamma C_{R\phib}^R)
\;,\;\;
\ee
\be
  C_{RL}^\phi 
= \frac{\dvp}{d_{\phi\phi}}
  (C_{L\phi}^R + \gamma C_{L\phi}^L)
\;,\;\;
  C_{RL}^\phib 
= \frac{\dvp}{d_{\phi\phi}}
  (C_{L\phib}^R + \gamma C_{L\phib}^L)
\;.\;\;
\ee

Finally, taking only $b$ chiral, we get
\be
  C_{R\phi}^\phib d_{\phib\phib} 
= d_{RR} C_{\phi\phib}^R  + d_{RL} C_{\phi\phib}^L 
\;,\;\;
  C_{L\phi}^\phib d_{\phib\phib} 
= d_{LR} C_{\phi\phib}^R  + d_{LL} C_{\phi\phib}^L 
\;,\;\;
\ee\be
  C_{R\phib}^\phi d_{\phi\phi} 
= d_{RR} C_{\phib\phi}^R  + d_{RL} C_{\phib\phi}^L 
\;,\;\;
  C_{L\phib}^\phi d_{\phi\phi} 
= d_{LR} C_{\phib\phi}^R  + d_{LL} C_{\phib\phi}^L 
\;.
\ee
Looking at the first of these, it becomes
\begin{align}
  C_{R\phi}^\phib 
&= \frac{1}{d_{\phib\phib} }
( d_{RR} C_{\phi\phib}^R  + d_{RL} C_{\phi\phib}^L  )
\nn\\
&= \frac{d_{\vp\vp}}{d_{\phi\phi}} ( C_{\phi\phib}^R  + \gamma C_{\phi\phib}^L )
\nn\\
&= \frac{d_{\vp\vp}}{d_{\phi\phi}} ( C_{\phi\phib}^R  + \gamma
C_{\phib\phi}^R )
\;.
\end{align}
Likewise we get
\begin{align}
  C_{L\phi}^\phib 
&= \frac{d_{\vp\vp}}{d_{\phi\phi}} 
   (\gamma C_{\phi\phib}^R  + C_{\phib\phi}^R) 
\;,\;\;
\\
  C_{R\phib}^\phi 
&= \frac{d_{\vp\vp}}{d_{\phi\phi}} 
   (C_{\phib\phi}^R  + \gamma C_{\phi\phib}^R) 
\;,\;\;
\\
  C_{L\phib}^\phi
&= \frac{d_{\vp\vp}}{d_{\phi\phi}}
    (\gamma C_{\phib\phi}^R  + C_{\phi\phib}^R) 
\;,
\end{align}
which also imply
\be
  C_{R\phib}^\phi =    C_{L\phi}^\phib 
\;,\;\;
  C_{L\phib}^\phi =    C_{R\phi}^\phib 
\;.
\ee

\subsubsection{Bulk field expectation operator product}

To find $C_{LR}^R$ we use the inner product matrix $d_{\alpha\beta}$
of defect fields $\vp_L$ and $\vp_R$ and cyclicity of the three point
constant
$C_{\alpha\beta\gamma}$ defined by
\be
 \langle \vp_\alpha(u,\bar u) \vp_\beta(v,\bar v) \vp_\gamma(w,\bar w) \rangle
= C_{\alpha\beta\gamma} \left( |u-v| |v-w| |v-w| \right)^{-2h}
\;.
\ee
Using $C_{\alpha\beta}^\gamma =
d^{\gamma\epsilon}C_{\alpha\beta\epsilon}$
and $C_{\alpha\beta\gamma} = C_{\gamma\beta\alpha}$ and the relations
\eqref{eq:319} and \eqref{eq:321}, we get

\begin{align}
  C_{LR}^R 
&= d^{RR} C_{LRR} + d^{RL} C_{LRL}
\nn\\
&= d^{RR} C_{RRL} + d^{RL} C_{LLR}
\nn\\
&= d^{RR} (d_{LL}C_{RR}^L + d_{LR} C_{RR}^R) + 
   d^{RL} (d_{RL}C_{LL}^L + d_{RR} C_{LL}^R)
\nn\\
&= (d^{RR} d_{LR} + d^{RL} d_{RL} )C_{\vp\vp}^\vp
\nn\\
&= (d^{RR} + d^{RL}) d_{RL} C_{\vp\vp}^\vp
\;.
\end{align}

With the inner-product matrix $d_{\alpha\beta} = \langle\vp_\alpha\vert\vp_\beta\rangle$,
\be
  d_{\alpha\beta} 
= \begin{pmatrix}
  d_{LL} & d_{LR} \\ d_{RL} & d_{RR} 
  \end{pmatrix}
= d_{\vp\vp}
  \begin{pmatrix}
  1 & \gamma \\ \gamma & 1
  \end{pmatrix}
\;,
\ee
and its inverse
\be
  d^{\alpha\beta} 
= \begin{pmatrix}
  d^{LL} & d^{LR} \\ d^{RL} & d^{RR} 
  \end{pmatrix}
= \frac{1}{d_{\vp\vp} (1 - \gamma^2)}
  \begin{pmatrix}
  1 & -\gamma \\ -\gamma & 1
  \end{pmatrix}
\;,
\ee
we obtain
\be
  C_{LR}^R 
%=  (d^{RR} + d^{RL}) d_{RL} C_{\vp\vp}^\vp
=  \frac{\gamma}{1 + \gamma} C_{\vp\vp}^\vp
\;.
\ee

Likewise, we find all four of these structure constants are equal,
\be
  C_{RL}^R = C_{LR}^L = C_{RL}^L =C_{LR}^R 
=  \frac{\gamma}{1 + \gamma} C_{\vp\vp}^\vp
\;.
\label{eq:352}
\ee

\subsubsection{Continuity of bulk fields}

\begin{figure}[htb]
\[
  \begin{tikzpicture}

\draw (-5,0) -- (-1,0);
\draw (1,0) -- (5,0);
\draw (7,0) -- (11,0);

\draw (-4.6,.1) -- (-4.5,0) -- (-4.6,-.1);
\draw (1.4,.1) -- (1.5,0) -- (1.4,-.1);
\draw (7.4,.1) -- (7.5,0) -- (7.4,-.1);

\filldraw (-3,0) circle (2pt);
\filldraw ( 3,0) circle (2pt);
\filldraw ( 9,0) circle (2pt);

\filldraw (-4,0) circle (2pt);
\filldraw ( 3,1) circle (2pt);
\filldraw (10,0) circle (2pt);

\draw[dashed] [->] (-3,1) arc (90:170:1);
\draw[dashed] [->] ( 9,1) arc (90:10:1);

\node at (-2.6,-.4) {$\Phi_a(x)$} ;
\node at ( 3,-.4) {$\Phi_a(x)$} ;
\node at ( 8.6,-.4) {$\Phi_a(x)$} ;

\node at (-4,-.4) {$\vp_L(x{-}1)$} ;
\node at ( 3,1.3) {$\vp(x+i)$} ;
\node at (10.4,-.4) {$\vp_L(x{+}1)$} ;

\node at (-3.6,-0.6) {$\underbrace{\phantom{fffffffffC_{La}^b}}$} ;
\node at ( 9.6,-0.6) {$\underbrace{\phantom{fffffffffC_{La}^b}}$} ;

\node at (-3.6,-1.2) {$C_{La}^b$} ;
\node at ( 9.6,-1.2) {$C_{aL}^b$} ;

  \end{tikzpicture}
\]
\caption{The relation between $C_{La}^b$ and $C_{aL}^b$ from
  continuity in the bulk.}
\label{fig:continuity}
\end{figure}
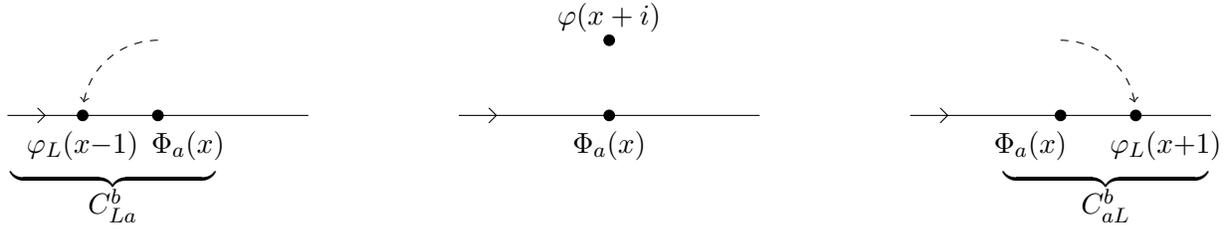

We can relate the structure constants $C_{a L}^b$ and $C_{L,a}^b$ by
moving the insertion point of the field $\vp_L$ from the right of the
field $a$ to the left through the bulk. 
If the defect is oriented along the $x$ axis in the plane, then the
field $\vp_L$ can be moved through the upper half plane, as in figure
\ref{fig:continuity}.

Likewise, we can relate  $C_{a R}^b$ and $C_{R,a}^b$ by moving the
field $\vp_R$ through the lower half plane.

Since the OPEs of the bulk field $\vp$£ and the defect field  $\vp_L$ with $\Phi_a$ are
\be
  \phi_a(u,\bar u) \vp(z,\bar z) 
= C_{a\vp}^b \Phi_b(u,\bar u) 
  (u-z)^{h_b-h_a-h} (\bar u-\bar z)^{\bar h_b - \bar h_a - h} + \ldots
\;,
\ee
\be
  \phi_a(u,\bar u) \vp_L(z,\bar z) 
= C_{aL}^b \Phi_b(u,\bar u) 
  \ob u-z \cb^{h_b-h_a-h} \ob \bar u-\bar z \cb^{\bar h_b - \bar h_a - h} + \ldots
\;,
\ee
\be
  \vp_L(z,\bar z) \phi_a(u,\bar u) 
= C_{La}^b \Phi_b(u,\bar u) 
  \ob z-u \cb^{h_b-h_a-h} \ob \bar z-\bar u \cb^{\bar h_b - \bar h_a - h} + \ldots
\;,
\ee
we get the relations
\begin{align} 
  C_{La}^b &= \exp( i \pi( h_b - \bar h_b - h_a + \bar h_a) ) C_{aL}^b
\;,\;\;
\\
  C_{Ra}^b &= \exp( -i \pi( h_b - \bar h_b - h_a + \bar h_a) ) C_{aR}^b
\;.
\end{align}
We again list the cases according to the number of chiral fields involved:

$\bullet$ No chiral fields: we find identities consistent with
equation \eqref{eq:352}
\be
 C_{LR}^R = C_{RL}^R
\;,\;\;
 C_{LR}^L = C_{RL}^L
\;.
\ee

$\bullet$ If $\Phi_b$ is chiral and $\Phi_a$ is not; with $\zeta=\exp(i\pi h)$:
\be
 C_{L\alpha}^\phi = \zeta C_{\alpha L}^\phi
\;,\;\;
 C_{L\alpha}^\phib = \zeta^{-1} C_{\alpha L}^\phib
\;,\;\;
 C_{R\alpha}^\phi = \zeta^{-1} C_{\alpha R}^\phi
\;,\;\;
 C_{R\alpha}^\phib = \zeta C_{\alpha R}^\phib
\;,\;\;
\ee
and hence
\be
 C_{LL}^\phi = C_{LL}^\phib = 
 C_{RR}^\phi = C_{RR}^\phib = 0
\;,\;\;
 C_{LR}^\phi = \zeta C_{RL}^\phi
\;,\;\;
 C_{LR}^\phib = \zeta^{-1} C_{RL}^\phib
\;.
\ee
where the first four structure constants were already found to be zero
in equations \eqref{eq:319} and \eqref{eq:321}.

$\bullet$ If $\Phi_a$ is chiral and $\Phi_b$ is not:
\be
 C_{L\phi}^L = \zeta^{-1} C_{\phi L}^L
\;,\;\;
 C_{L\phi}^R = \zeta^{-1} C_{\phi L}^R
\;,\;\;
 C_{L\phib}^L = \zeta C_{\phib L}^L
\;,\;\;
 C_{L\phib}^R = \zeta C_{\phib L}^R
\;,\;\;
\ee
\be
 C_{R\phi}^L = \zeta C_{\phi R}^L
\;,\;\;
 C_{R\phi}^R = \zeta C_{\phi R}^R
\;,\;\;
 C_{R\phib}^L = \zeta^{-1} C_{\phib R}^L
\;,\;\;
 C_{R\phib}^R = \zeta^{-1} C_{\phib R}^R
\;,\;\;
\ee
\be
 C_{LR}^\phi = \zeta C_{RL}^\phi
\;,\;\;
 C_{LR}^\phib = \zeta^{-1} C_{RL}^\phib
\;.\;\;
\ee

$\bullet$ If both $\Phi_a$ and $\Phi_b$ are chiral:
\be
 C_{L\phi}^\phib = \zeta^{-2} C_{\phi L}^\phib
\;,\;\;
 C_{L\phib}^\phi =  \zeta^2 C_{\phib L}^\phi
\;,\;\;
 C_{R\phi}^\phib = \zeta^{2} C_{\phi R}^\phib
\;,\;\;
 C_{R\phib}^\phi =  \zeta^{-2} C_{\phib R}^\phi
\;,\;\;
\ee

\subsection{Unknown constants}

We summarise the results so far, distinguishing the structure
constants by the number of chiral fields they involve.

\subsubsection{No chiral fields}
These are all known in terms of the bulk field data:
\begin{align}
   d_{RR} = d_{LL}
= d_{\vp\vp}
\;,&\;\;\;\;
   d_{LR} = d_{RL}
= \gamma \,d_{\vp\vp}
\;,
\\[3mm]
   C_{LL}^L = C_{RR}^R 
= C_{\vp\vp}^\vp
\;,&\;\;\;\;
   C_{LL}^R = C_{RR}^L 
= 0
\;,
\end{align}
\be
   C_{LR}^R = C_{LR}^L = C_{RL}^R = C_{RL}^L 
= \frac{\gamma}{1 + \gamma}\, C_{\vp\vp}^\vp
\;.
\ee
\subsubsection{Three chiral fields}
These are also all known in terms of the boundary field theory data \cite{runkel}:
\be
   C_{\phib\phib}^\phib
= C_{\phi\phi}^\phi
\;,\;\;\;\;
   d_{\phib\phib}
= d_{\phi\phi}
\;,
\ee
\be
   C_{\phib\phib}^\phi = 
   C_{\phi\phi}^\phib = 
   C_{\phi\phib}^\phib = 
   C_{\phib\phi}^\phib = 
   C_{\phi\phib}^\phi = 
   C_{\phib\phi}^\phi  
= 0
\;.
\ee

\subsubsection{Two chiral fields}
The 24 structure constants involving two chiral fields
can be written in terms of just two of these,
which we can take to be
\be
 C_{\phib\phi}^L\;,\;\;
\hbox{and}\;\;
 C_{\phi\phib}^L
\;.
\ee
Listing the remaining 22 structure constants:
\begin{align} 
 C_{\phi\phib}^R 
&= C_{\phib\phi}^L
\;,\;\;&
 C_{\phib\phi}^R 
&= C_{\phi\phib}^L
\;,\;\;\\
 C_{R \phi}^\phib
&= 
 \frac{\dvp}{d_{\phi\phi}} ( C_{\phib\phi}^L + \gamma C_{\phi\phib}^L)
\;,\;\;&
  C_{L\phi}^\phib 
&= 
 \frac{\dvp}{d_{\phi\phi}} (\gamma C_{\phib\phi}^L  + C_{\phi\phib}^L) 
\;,\;\;\\
  C_{R\phib}^\phi 
&= 
  \frac{\dvp}{d_{\phi\phi}} (C_{\phi\phib}^L  + \gamma C_{\phib\phi}^L) 
\;,\;\;&
  C_{L\phib}^\phi
&= 
  \frac{\dvp}{d_{\phi\phi}} (\gamma C_{\phi\phib}^L  + C_{\phib\phi}^L) 
\;,\;\;\\
 C_{\phi R}^\phib
&= \zeta^{-2}
 \frac{\dvp}{d_{\phi\phi}} ( C_{\phib\phi}^L + \gamma C_{\phi\phib}^L)
\;,\;\;&
  C_{\phi L}^\phib 
&= \zeta^2
 \frac{\dvp}{d_{\phi\phi}} (\gamma C_{\phib\phi}^L  + C_{\phi\phib}^L) 
\;,\;\;\\
  C_{\phib R}^\phi 
&= \zeta^2
  \frac{\dvp}{d_{\phi\phi}} (C_{\phi\phib}^L  + \gamma C_{\phib\phi}^L) 
\;,\;\;&\
  C_{\phib L}^\phi
&= \zeta^{-2}
  \frac{\dvp}{d_{\phi\phi}} (\gamma C_{\phi\phib}^L  + C_{\phib\phi}^L) 
\;,
\end{align}
\begin{align}
 C_{\phi\phi}^R = 
 C_{\phi\phi}^L = 
 C_{\phib\phib}^R = 
 C_{\phib\phib}^L 
&= 0
\;,
\\
 C_{R\phi}^\phi = 
 C_{L\phi}^\phi = 
 C_{R\phib}^\phib = 
 C_{L\phib}^\phib
&= 0
\;,
\\ 
 C_{\phi R}^\phi = 
 C_{\phi L}^\phi = 
 C_{\phib R}^\phib = 
 C_{\phib L}^\phib
&= 0
\;.
\end{align}

It will be convenient to introduce $\rhoo$ and $\Gamma$ to parametrise
$ C_{\phi\phib}^L $ and $ C_{\phib\phi}^L$ as
\be
 C_{\phi\phib}^L = \rhoo \Gamma
\;,\;\;
 C_{\phib\phi}^L = \rhoo^{-1} \Gamma
\;,\;\;
C_{\phi\phib}^L = \rhoo^2 C_{\phib\phi}^L
\;.
\ee
It will turn out that $\Gamma$ is real and non-negative and $\rhoo$ is
a pure phase. 
%>>>>>>>>>>>>>>>>>>>
We note that these two structure constants can be found from the
results in \cite{IR10} - they are related to $C_s$ defined in
\cite{IR10}:eqn (2.19).
%<<<<<<<<<<<<<<<<<<

\subsubsection{One chiral field}

The twenty-four structure constants involving just one chiral
field can, using the previous identities, be written in terms of just
four:
\be
C_{\phi L}^R
\;,\;\;
C_{\phib L}^R
\;,\;\;
C_{\phi R}^L
\;,\;\;
C_{\phib R}^L
\;.
\ee
We list the remaining twenty constants here for convenience:
\begin{align}
C_{L\phi}^R 
&= \zeta^{-1} C_{\phi L}^R
\;,\;\; &
C_{L\phib}^R 
&= \zeta C_{\phib L}^R
\;,\\
C_{R\phi}^L
&= \zeta^{-1} C_{\phi R}^L
\;,\;\; &
C_{R\phib}^L 
&= \zeta C_{\phib R}^L
\;,\\
C_{\phi L}^L
&= -\gamma C_{\phi L}^R
\;,\;\;&
C_{\phi R}^R
&= -\gamma C_{\phi R}^L
\;,\\
C_{\phib L}^L
&= -\gamma C_{\phib L}^R
\;,\;\;&
C_{\phib R}^R
&= -\gamma C_{\phib R}^L
\;,\\
C_{L\phi}^L
&= \zeta^{-1} C_{\phi L}^L
= -\gamma \zeta^{-1}\, C_{\phi L}^R
\;,\;\;&
C_{L\phib}^L
&= \zeta C_{\phib L}^L
= -\gamma \zeta\, C_{\phib L}^R
\;,\\
C_{R\phi}^R
&= \zeta^{-1} C_{\phi R}^R
= -\gamma \zeta^{-1}\, C_{\phi R}^L
\;,\;\;&
C_{R\phib}^R
&= \zeta C_{\phib R}^R
= -\gamma \zeta\, C_{\phib R}^L
\;,\\
C_{LR}^\phi
&= \frac{1-\gamma^2}{\zeta} \frac{\dvp}{d_{\phi\phi}} C_{\phi R}^L
\;,\;\;&
C_{LR}^\phib
&= ({1-\gamma^2}) \frac{\dvp}{d_{\phi\phi}} C_{\phib L}^R
\;,\\
C_{RL}^\phi 
&=  \frac{1-\gamma^2}{\zeta^2} \frac{\dvp}{d_{\phi\phi}} C_{\phi R}^L
\;,\;\;&
C_{RL}^\phib 
&= \zeta\,({1-\gamma^2}) \frac{\dvp}{d_{\phi\phi}} C_{\phib L}^R
\;,
\end{align}
\be
C_{LL}^\phi = 
C_{LL}^\phib = 
C_{RR}^\phi = 
C_{RR}^\phib 
= 0
\;.\ee

%\newpage

\subsection{The four-point function sewing constraints}
\label{fpf}

We will use crossing relations for four point correlation functions to
find sewing constraints that will enable us to determined the
remaining six structure constants 
$\{
C_{\phib\phi}^L,C_{\phi\phib}^L,
C_{\phi L}^R,
C_{\phib L}^R,
C_{\phi R}^L,
C_{\phib R}^L
\}$.

%From figure \ref{crossing}, 
The four-point function 
$\langle \Phi_a \Phi_b \Phi_c \Phi_d \rangle$ of fields on a defect
can be expressed
in terms of conformal blocks in two different ways, as illustrated in
figure \ref{crossing}

\begin{figure}[htb]
\[
\begin{tikzpicture}

\draw (-2,2) circle (1) ;
\draw (-2,-2) circle (1cm) ;

\draw (-3.1,1.9) -- (-3,2) -- (-2.9,1.9);
\draw[dashed] (-2.7,2.7) -- (-2,2.4) -- (-2,1.6) -- (-2.7,1.3);
\draw[dashed] (-1.3,2.7) -- (-2,2.4); 
\draw[dashed] (-2,1.6) -- (-1.3,1.3); 
\node at (-2.9,2.9) {$a$} ;
\node at (-1.1,2.9) {$b$} ;
\node at (-1.1,1.1) {$c$} ;
\node at (-2.9,1.1) {$d$} ;
\node at (-2,2.6) {$e$} ;
\node at (-2,1.35) {$f$} ;
\filldraw (-2.7,2.7) circle (2pt);
\filldraw (-2.7,1.3) circle (2pt);
\filldraw (-1.3,2.7) circle (2pt);
\filldraw (-1.3,1.3) circle (2pt);

\draw (-3.1,-2.1) -- (-3,-2) -- (-2.9,-2.1);
\draw[dashed] (-2.7,-2.7) -- (-2.4,-2) -- (-1.6,-2) -- (-1.3,-1.3);
\draw[dashed] (-2.7,-1.3) -- (-2.4,-2); 
\draw[dashed] (-1.6,-2) -- (-1.3,-2.7); 
\node at (-2.9,-2.9) {$d$} ;
\node at (-1.1,-2.9) {$c$} ;
\node at (-1.1,-1.1) {$b$} ;
\node at (-2.9,-1.1) {$a$} ;
\node at (-2.6,-2) {$k$} ;
\node at (-1.4,-2) {$g$} ;
\filldraw (-2.7,-2.7) circle (2pt);
\filldraw (-2.7,-1.3) circle (2pt);
\filldraw (-1.3,-2.7) circle (2pt);
\filldraw (-1.3,-1.3) circle (2pt);

\node[right] at (0,2) {$\ds = \sum_{ef} C_{ab}^e C_{cd}^f d_{ef} 
\left(\raisebox{-2mm}{\cbI {2300}{h_d}{h_a}{h_b}{h_c}{h_e}{}{}{}}\right)
\left(\raisebox{-2mm}{\cbI {2300}{\bar h_d}{\bar h_a}{\bar h_b}{\bar h_c}{\bar h_e}{}{}{}}\right)^*
\delta_{h_e,h_f} \delta_{\bar h_e,\bar h_f}
$};

\node[right] at (0,-2) {$\ds = \sum_{kg} C_{bc}^g C_{da}^k d_{gk} 
\left(\raisebox{-2mm}{\cbb {1500}{h_d}{h_a}{h_k}{h_b}{h_c}{}{}{}}\right)
\left(\raisebox{-2mm}{\cbb {1500}{\bar h_d}{\bar h_a}{\bar h_k}{\bar h_b}{\bar h_c}{}{}{}}\right)^*
\delta_{h_k,h_g} \delta_{\bar h_k,\bar h_g}
$};

\end{tikzpicture}
\]
\caption{Two ways of calculating a four-point defect field correlation function}
\label{crossing}
\end{figure}

\blank{
\begin{figure}[htb]
\[
  \begin{picture}(450,240)
  \put(0,0){\scalebox{.8}{\includegraphics{Crossing_cropped.eps}}}
  \end{picture}
\]
\caption{Two ways of calculating a four-point defect field correlation function}
\label{crossing}
\end{figure}
}

\blank{
\begin{align}
& \sum_{e,f} C_{ab}^e C_{cd}^f d_{ef}
 G(h_a, h_b, h_c, h_d | h_e)(x)
 \bar G(\bar h_a, \bar h_b, \bar h_c, \bar h_d | \bar h_e)(\bar x)
 \delta_{h_e,f_f} \delta{\bar h_e,\bar h_f}
\nn\\
&= 
 \sum_{g,k} C_{bc}^g C_{da}^k d_{gk}
 G(h_b, h_c, h_d, h_a | h_g)(1-x)
 \bar G(\bar h_b, \bar h_c, \bar h_d, \bar h_a | \bar h_g)(1 - \bar x)
 \delta_{h_g,f_k} \delta{\bar h_g,\bar h_k}
\;.
\end{align}
}

The conformal blocks are functions which satisfy the crossing
relations \cite{runkel} 
\be
\raisebox{-5mm}{\mbox{
$\cbb{1500}ijpkl{}{}
\;\;
$}}
= \sum_q
F\left[{\scriptstyle
\begin{matrix}j & k \\ i & l \end{matrix}}\right]_{pq}
\raisebox{-2.5mm}{\mbox{
$\cbI{2300}ijklq{1}{z}
\;\;
$}}
\label{eq:Fdef}
\ee
where the F-matrices are known constants, again given explicitly in
\cite{runkel}. Substituting \eqref{eq:Fdef} into the expressions in
figure \ref{crossing} leads to further sewing constraints that the
structure constants must satisfy. 

The simplest relations arise when there is only a single channel in
both diagrams, i.e. the sum is over a single pair of weights
$(h_e,\bar h_e)$  and a single pair of weights $(h_g,\bar h_g)$.
Note that since the space of fields with weights $(h,h)$ is
two-dimensional, this does not mean that the 
OPE has to include only a single field. 
In all the cases where there is only a single channel, the $F$-matrix
is just the number 1 and so the sewing constraints become just
\be
  \sum_{e,f} C_{ab}^e C_{cd}^f d_{ef}
= \sum_{g,k} C_{bc}^g C_{da}^k d_{gk}
\;.
\ee

We now list all the non-zero cases in which the fields $a,b,c$ and $d$ are
taken from $\{\phi,\phib,\vp_\alpha\}$ and for which there is only a
single intermediate channel in both diagrams, and state the
corresponding equations. 
We will in fact only use the first eight of these, where there is at
most one field of weights $(h,h)$ but we list them all for completeness.
The eight we use are:
\begin{align}
\fppA{8000}{\phi}{\phi}{\phib}{\phib}
& 
 d_{\phi\phi}\, d_{\phib\phib}
= 
 \sum_{\alpha,\beta} 
 C_{\phi\phib}^\alpha C_{\phib\phi}^\beta d_{\alpha\beta}
\label{i}
\\
\fppA{8000}{\phi}{\phib}{\phi}{\phib}
& 
 \sum_{\alpha,\beta} 
 C_{\phi\phib}^\alpha C_{\phi\phib}^\beta d_{\alpha\beta}
= 
 \sum_{\alpha,\beta} 
 C_{\phib\phi}^\alpha C_{\phib\phi}^\beta d_{\alpha\beta}
\label{ii}
\displaybreak[0]\\
\\
\fppA{8000}{\alpha}{\phib}{\phi}{\phi}
& 
 C_{\alpha\phib}^\phi C_{\phi\phi}^\phi d_{\phi\phi}
= 
 \sum_{\beta,\gamma} 
 C_{\phib\phi}^\beta C_{\phi\alpha}^\gamma d_{\beta\gamma}
\label{iii}
\displaybreak[0]\\
\fppA{8000}{\alpha}{\phi}{\phib}{\phi}
& 
 \sum_{\beta\gamma}
 C_{\alpha\phi}^\beta C_{\phib\phi}^\gamma d_{\beta\gamma}
= 
 \sum_{\beta,\gamma} 
 C_{\phi\phib}^\beta C_{\phi\alpha}^\gamma d_{\beta\gamma}
\label{iv}
\displaybreak[0]\\
\fppA{8000}{\alpha}{\phi}{\phi}{\phib}
& 
 \sum_{\beta\gamma}
 C_{\alpha\phi}^\beta C_{\phi\phib}^\gamma d_{\beta\gamma}
= 
 C_{\phi\phi}^\phi C_{\phib\alpha}^\phi d_{\phi\phi}
\label{v}
\displaybreak[0]\\
\fppA{8000}{\alpha}{\phi}{\phib}{\phib}
& 
 C_{\alpha\phi}^\phib C_{\phib\phib}^\phib d_{\phib\phib}
= 
 \sum_{\beta,\gamma} 
 C_{\phi\phib}^\beta C_{\phib\alpha}^\gamma d_{\beta\gamma}
\label{vi}
\displaybreak[0]\\
\fppA{8000}{\alpha}{\phib}{\phi}{\phib}
& 
 \sum_{\beta\gamma}
 C_{\alpha\phib}^\beta C_{\phi\phib}^\gamma d_{\beta\gamma}
= 
 \sum_{\beta,\gamma} 
 C_{\phib\phi}^\beta C_{\phib\alpha}^\gamma d_{\beta\gamma}
\label{vii}
\displaybreak[0]\\
\fppA{8000}{\alpha}{\phib}{\phib}{\phi}
& 
 \sum_{\beta\gamma}
 C_{\alpha\phib}^\beta C_{\phib\phi}^\gamma d_{\beta\gamma}
= 
 C_{\phib\phib}^\phib C_{\phi\alpha}^\phib d_{\phib\phib}
\label{viii}
\\\nonumber
\end{align}

The remaining three which include two fields of type
$\vp_\alpha$ but still only have a single intermediate channel are:
\begin{align}
\fppA{8000}{\alpha}{\beta}{\phi}{\phib}
& 
 \sum_{\gamma\epsilon}
 C_{\alpha\beta}^\gamma C_{\phi\phib}^\epsilon d_{\gamma\epsilon}
= 
 \sum_{\gamma\epsilon}
 C_{\beta\phi}^\gamma C_{\phib\alpha}^\epsilon d_{\gamma\epsilon}
\label{ix}
\displaybreak[0]\\
\\
\fppA{8000}{\alpha}{\beta}{\phib}{\phi}
& 
 \sum_{\gamma\epsilon}
 C_{\alpha\beta}^\gamma C_{\phib\phi}^\epsilon d_{\gamma\epsilon}
= 
 \sum_{\gamma\epsilon}
 C_{\beta\phib}^\gamma C_{\phi\alpha}^\epsilon d_{\gamma\epsilon}
\label{x}
\displaybreak[0]\\
\\
\fppA{8000}{\alpha}{\phi}{\beta}{\phib}
& 
 \sum_{\gamma\epsilon}
 C_{\alpha\phi}^\gamma C_{\beta\phib}^\epsilon d_{\gamma\epsilon}
= 
 \sum_{\gamma\epsilon}
 C_{\phi\beta}^\gamma C_{\phib\alpha}^\epsilon d_{\gamma\epsilon}
\label{xi}
\blank{
\\
\fppA{8000}LLL{\phi}
& 
 \sum_{\alpha}
 C_{LL}^L C_{L\phi}^\alpha d_{L\alpha}
= 
 \sum_{\alpha}
 C_{LL}^L C_{\phi L}^\alpha d_{L\alpha}
\label{xii}
\\
\fppA{8000}RRR{\phi}
& 
 \sum_{\alpha}
 C_{RR}^R C_{R\phi}^\alpha d_{R\alpha}
= 
 \sum_{\alpha}
 C_{RR}^R C_{\phi R}^\alpha d_{R\alpha}
\label{xiii}
\\
\fppA{8000}LLL{\phib}
& 
 \sum_{\alpha}
 C_{LL}^L C_{L\phib}^\alpha d_{L\alpha}
= 
 \sum_{\alpha}
 C_{LL}^L C_{\phib L}^\alpha d_{L\alpha}
\label{xiv}
\\
\fppA{8000}RRR{\phib}
& 
 \sum_{\alpha}
 C_{RR}^R C_{R\phib}^\alpha d_{R\alpha}
= 
 \sum_{\alpha}
 C_{RR}^R C_{\phib R}^\alpha d_{R\alpha}
\label{xv}
}
\\\nonumber
\end{align}

%\newpage
\subsection{Analysis of the sewing constraints}

We need to use only the first eight relations. We consider these in
turn: 

$\bullet$ {Equation \eqref{i}}

Written out in full, this is
\begin{align}
  d_{\phi\phi} d_{\phib\phib}
&=
  C_{\phi\phib}^L C_{\phib\phi}^L d_{LL}
+ C_{\phi\phib}^L C_{\phib\phi}^R d_{LR}
+ C_{\phi\phib}^R C_{\phib\phi}^L d_{RL}
+ C_{\phi\phib}^R C_{\phib\phi}^R d_{RR}
\;.
\end{align} 
Using 
$C_{\phi\phib}^L = C_{\phib\phi}^R = \rhoo \Gamma$
and 
$C_{\phi\phib}^R = C_{\phib\phi}^L = \rhoo^{-1} \Gamma$,
together with $d_{LR} = d_{RL} = \gamma \dvp$, 
and $d_{\phi\phi} = d_{\phib\phib}$, this becomes  
\be
  \frac{d_{\phi\phi}^2}{\dvp}
= \Gamma^2 ( 2 + \gamma\rhoo^2 + \gamma\rhoo^{-2})
\;,
\ee
or
\be
\Gamma = 
  \sqrt{
\frac{d_{\phi\phi}^2}{\dvp\,( 2 + \gamma\rhoo^2 + \gamma\rhoo^{-2})}
}
\;.
\ee

$\bullet$ {Equation \eqref{ii}}

This is
\begin{align}
 & C_{\phi\phib}^L C_{\phi\phib}^L d_{LL}
+ C_{\phi\phib}^L C_{\phi\phib}^R d_{LR}
+ C_{\phi\phib}^R C_{\phi\phib}^L d_{RL}
+ C_{\phi\phib}^R C_{\phi\phib}^R d_{RR}
\nn\\
&=
  C_{\phib\phi}^L C_{\phib\phi}^L d_{LL}
+ C_{\phib\phi}^L C_{\phib\phi}^R d_{LR}
+ C_{\phib\phi}^R C_{\phib\phi}^L d_{RL}
+ C_{\phib\phi}^R C_{\phib\phi}^R d_{RR}
\;,
\end{align}
which is satisfied identically

$\bullet$ {Equation \eqref{iii}}

This leads to two equations: for $\alpha=L$:
\be
  C_{L\phib}^\phi C_{\phi\phi}^\phi d_{\phi\phi}
= C_{\phib\phi}^L C_{\phi L}^L d_{LL}
+ C_{\phib\phi}^L C_{\phi L}^R d_{LR}
+ C_{\phib\phi}^R C_{\phi L}^L d_{RL}
+ C_{\phib\phi}^R C_{\phi L}^R d_{RR}
\;,
\ee
and for $\alpha=R$:
\be
  C_{R\phib}^\phi C_{\phi\phi}^\phi d_{\phi\phi}
= C_{\phib\phi}^L C_{\phi R}^L d_{LL}
+ C_{\phib\phi}^L C_{\phi R}^R d_{LR}
+ C_{\phib\phi}^R C_{\phi R}^L d_{RL}
+ C_{\phib\phi}^R C_{\phi R}^R d_{RR}
\ee
The first equation becomes:
\be
  (\gamma C_{\phi\phib}^L  + C_{\phib\phi}^L) 
  C_{\phi\phi}^\phi 
= C_{\phi L}^R C_{\phi\phib}^L
  \left(
  1 - \gamma^2
 \right)
\;,
\ee
or
\be
 C_{\phi L}^R = \frac{1 + \rhoo^2\gamma}{\rhoo^2(1 - \gamma^2)} 
               C_{\phi\phi}^\phi
\;.
\ee
The second equation implies
\be
 C_{\phi R}^L = \frac{\rhoo^2 + \gamma}{(1 - \gamma^2)} 
               C_{\phi\phi}^\phi
\;.
\ee

$\bullet$ {Equation \eqref{iv}}

These two equations imply
\be
  \rhoo^2 = \zeta = \exp(i \pi h)
\;.
\ee
(We will not need to fix the sign of $\rhoo$ as only $\rhoo^2$ appears
in our final answers)

$\bullet$ {Equation \eqref{v}}

These equations imply (for $\alpha=L$)
\be
  C_{\phi L}^R 
= \frac{ 1 + \rhoo^2\gamma}{\zeta(1 - \gamma^2)} C_{\phi\phi}^\phi
\;,
\ee
and (for $\alpha=R$)
\be
  C_{\phi R}^L
= \frac{\zeta^2}{\rhoo}\frac{ \gamma + \rhoo^2}{(1 - \gamma^2)} C_{\phi\phi}^\phi
\;,
\ee
which are consistent with the results so far.

$\bullet$ {Equation \eqref{vi}}

These two equations lead to
($\alpha = L$):
\be
 C_{\phib L}^R = \frac{\gamma + \rhoo^2}{1 - \gamma^2}
 C_{\phi\phi}^\phi
\;,
\ee
and (with $\alpha = R$)
\be
 C_{\phib R}^L = \frac{1 + \gamma \rhoo^2}{\rhoo^2(1 - \gamma^2)}
 C_{\phi\phi}^\phi
\;.
\ee
Together, these imply
\be
 C_{\phib L}^R = C_{\phi R}^L
\;\;\;\;\hbox{and}\;\;\;\;
 C_{\phib R}^L = C_{\phi L}^R
\;.
\ee

This completes the derivation of the structure constants.
They agree with the specific case in \cite{bhw} (apart from a typo in
\cite{bhw}, where it should $\rho=\exp(i\pi/10)$).

The remaining crossing relations 
\eqref{ix} -- \eqref{xi}
are not needed for the derivation
of the structure constants but we have checked that they hold.

%\newpage
\section{The integrals}
\label{ints}

We want to calculate the leading term in the expansion
\eqref{eq:T1Tb1}, that is 
\be
I 
= 
\frac{1}{4}
  \lambda^2\bar \lambda^2\int \rd x\, \rd x'\, \rd y\, \rd y'
  \vev{  T(iY)\,\olT(iY)\,  \phi(x) \phi(x') \bar\phi(y)  \bar\phi(y') }
\;.
\label{eq:Idef}
\ee

The correlation function has the same functional form whatever the
order of the fields, but a different constant depending on the order of
the insertions.
We can restrict to $x<x'$ and $y<y'$ to get 
\be
I = 
(\lambda\bar\lambda)^2 
\;
\left\langle
\;
T(i) \bar T(i) 
\int_{x<x'\,,\,y<y'} \rd x\, \rd x'\, \rd y\, \rd y'\, \phi(x) \phi(x') \phib(y) \phib(y')
\;
\right\rangle_{D_{r2}}
\;.
\ee
This correlation function is 
\be
\left\langle
\;
T(i)\, \bar T(i) 
\, \phi(x) \phi(x') \phib(y) \phib(y')
\;
\right\rangle
= \Delta\, h^2\, \frac{(x'-x)^{2-2h}\,(y'-y)^{2-2h}}
               {(i-x)^2(i-x')^2(i+y)^2(i+y')^2 } 
\;,
\label{eq:DeltaDef}
\ee
where the constant $\Delta$ depends on the order of the field
insertions as in table \ref{delta12}

\begin{table}[htb]
\[
\begin{array}{c|c|c}
\hbox{Integration region}
&
\hbox{Order of fields}
&
\hbox{Value of $\Delta$}
\\ \hline
x<x'<y<y' & \phi\phi\phib\phib & \Delta_1 \\
x<y<x'<y' & \phi\phib\phi\phib & \Delta_2 \\
x<y<y'<x' & \phi\phib\phib\phi & \Delta_1 \\
y<x<x'<y' & \phib\phi\phi\phib & \Delta_1 \\
y<x<y'<x' & \phib\phi\phib\phi & \Delta_2 \\
y<y'<x<x' & \phib\phib\phi\phi & \Delta_1
\end{array}
\]
\caption{The coefficient in the four-point function \eqref{eq:DeltaDef}}
\label{delta12}
\end{table}

The values $\Delta_i$ are
\begin{align}
\Delta_1 =& d_{\phi\phi}\, d_{\phib\phib} 
= (d_{\phi\phi})^2
\;,
\\
 \Delta_2 =& d_{\alpha\beta}\, C_{\phi\phib}^\alpha
  C_{\phi\phib}^\beta 
=
  (d_{\phi\phi})^2 
  \frac{ 2 \gamma + \rhoo^2 + \rhoo^{-2}}
       { 2 + \gamma\rhoo^2 + \gamma\rhoo^{-2}}
\;.
\end{align}

We only need to evaluate three of these integrations, the other three
being given by complex conjugation. Furthermore, we only need the
leading order 
term in $y$ in the correlation function,
\be
\left\langle
\;
T(i) \bar T(i) 
\phi(x) \phi(x') \phib(y) \phib(y')
\;
\right\rangle_{D_{r2}}
=
\frac{\Delta}{(i-x)^2(i-x')^2(i+y)^2(i+y')^2}
+ O(y)
\;.
\ee
The results are given in table \ref{tab:ints}.
\begin{table}[htb]
\[
\begin{array}{c|c|c}
\hbox{Integration region}
&
\hbox{Order of fields}
&
\hbox{Value of the integral}
\\ \hline
x<x'<y<y' & \phi\phi\phib\phib & -\frac{3\pi i}{16}\Delta_1 \\
x<y<x'<y' & \phi\phib\phi\phib & -\frac{\pi^2 + 3\pi i}{8}\Delta_2 \\
x<y<y'<x' & \phi\phib\phib\phi & \frac{\pi^2}{8}\Delta_1 \\
y<x<x'<y' & \phib\phi\phi\phib & \frac{\pi^2}{8}\Delta_1 \\
y<x<y'<x' & \phib\phi\phib\phi & -\frac{\pi^2 - 3\pi i}{8}\Delta_2 \\
y<y'<x<x' & \phib\phib\phi\phi & \frac{3\pi i}{16}\Delta_1
\end{array}
\]
\caption{The integrals}
\label{tab:ints}
\end{table}
Adding all six together, we get
\begin{align}
I =& (\lambda\bar\lambda)^2
\int_{-\infty}^\infty \rd x\, \rd x'\, \rd y\, \rd y'\, 
\left\langle
\;
T(i) \bar T(i) 
\phi(x) \phi(x') \phib(y) \phib(y')
\;
\right\rangle_{D_{r2}}
\nonumber\\
=& 
(\lambda\bar\lambda)^2
\left[\frac{\pi^2}{4}(\Delta_1 - \Delta_2) + O(y) \right]
\nonumber\\
=& 
\frac{\pi^2}4
\;
(\lambda\bar\lambda)^2 
\;
(d_{\phi\phi})^2
\,
\left[
1 - 
  \left[\frac{ 2 \gamma + \rhoo^2 + \rhoo^{-2}}
       { 2 + \gamma\rhoo^2 + \gamma\rhoo^{-2}}\right]
+ O(y)\right]
\;.
\end{align}

\section{The value of the reflection coefficient for the defect $C$}
\label{ans}

We now put the various terms together to find the value of $\cR$ at
the fixed point $(\lambda^*,\lambda^*)$,
\be
\cR
= \frac{\langle T^1 \olT{}^1 + T^2 \olT{}^2\rangle}
       {\langle (T^1 + \olT{}^2)( \olT{}^1 + T^2)\rangle}
\;.
\ee
The leading term in the numerator is $2I$ and leading term in the
denominator is $c/16$.

We first give the expansion in $y=1-h$ of the various constants.
With $h=2t-1$ we get $t=1-y/2$ and so
\begin{align}
%   h
%&= 1 - y
%\\
%   t
%&= 1 - y/2
%\\
   \rhoo^2 = \zeta 
= \exp(i \pi h) &= -1 + O(y)
\;,
\\
   \gamma 
= 2\cos(2\pi t)-1 &= 1 + O(y^2)
\;,
\\
\frac{ (C_{\phi\phi}^\phi)^2}{d_{\phi\phi}}
&=\frac 83 + O(y)
\;.
\end{align}
At the fixed point,
\begin{align}
I =&
\frac{\pi^2}4
\;
(\lambda^*)^4 
\;
(d_{\phi\phi})^2
\,
\left[
1 - 
  \left[\frac{ 2 \gamma + \rhoo^2 + \rhoo^{-2}}
       { 2 + \gamma\rhoo^2 + \gamma\rhoo^{-2}}\right]
+ O(y)\right]
\nonumber\\
=&
\frac{9 \pi^2 y^4}{256}  + O(y^5)
\;,
\end{align}
and with $c=1 + O(y)$, we find 
\be
\cR = 
\frac{2 \frac{9 \pi^2 y^4}{256}  + O(y^5)}
     {1/16 + O(y)}
=
\frac{9\pi^2 y^4}{8} + O(y^5)
\;.
\ee

We can now calculate this for the tri-critical Ising model.
In this case, $h=3/5$, $y=2/5$ and we are far from the small $y$
regime, but we calculate the leading correction and get
\be
{\cal R} \sim  
\frac{18\pi^2}{625}
= 0.284..
\ee

This can be compared with the values in \cite{mw}, which are
\be
\frac{\sqrt{3}-1}2 = 0.366..
\;\;\;\;\hbox{and}\;\;\;\;
\frac{3- \sqrt{3}}2 = 0.633..
\;.
\ee

%\newpage

\section{Conclusions}
\label{concs}

We have calculated the leading term in the perturbative expansion of
the reflection coefficient for the defect of type \rt\ in a minimal
model.
It is believed that a non-trivial conformal defect can be found as a
perturbative fixed point of the renormalisation group equations.
We have recently found new non-trivial conformal defects in the
tri-critical Ising model \cite{mw} and it is possible that these are
related to the conformal defects found by perturbation theory.
We have checked, and the value of $\cR$ is close enough not to rule
this out. It would of course be good to extend this calculation to
next-to-leading order where there are UV divergences to be regulated,
but so far we have not yet managed this.

We have also calculated defect structure constants for various fields
on defects of type \rt\ extending the results of \cite{bhw}. These
results are not complete - they do not include all fields, and use
special properties of the \rt\ defect, but it would be good to check
that these constants in fact agree with the general results of
\cite{RCFTIV} where the same constants were constructed using
topological field theory methods.

We would like to thank I.~Runkel, C.~Schmidt-Colinet and E.~Brehm for
discussions on defects and their properties and for comments on the
manuscript. 

\appendix
\section{The Virasoro Minimal Models}
\label{app:ir}

The Virasoro minimal models occur for $c\equiv c(p,q)$ where 
$p,q$ are coprime positive integers greater than 1.
It is useful to define $t=p/q$. $c$ is given by
\be
c(p,q) = 13 - 6 t - 6 /t
\;.
\ee
There are $(p-1)(q-1)/2$ minimal representations labelled by integers
$(r,s)$ with $1\leq r <p$, $1\leq s<q$ with conformal
weights
\be
  h_{r,s} 
= \frac{ (r q-sp)^2 - (p-q)^2}{4pq}
= \frac{r^2-1}{4t} + \frac{s^2-1}4 t - \frac{rs - 1}2
\;.
\ee
The modular S-matrix is
\be
 S_{(r,s),(r',s')} = 
(-1)^{1 + r s' + r' s} \sqrt{\frac{8}{pq}}
 \sin(\pi r r' / t) \sin(\pi s s' t)
\;.
\ee

%\newpage
%\bibliographystyle{JHEP}
%\bibliography{draft}

\begin{thebibliography}{10}

%\bibitem{YBK}

%1
\bibitem{DMS}
G.~Delfino, G.~Mussardo and P.~Simonetti,
\emph{Statistical Models with a Line of Defect},
\doi{10.1016/0370-2693(94)90439-1}{Phys.~Lett.~B328 (1994) 123-129}
[\arxiv{hep-th/9403049}{hep-th/9403049}]

%2
\bibitem{QRW}
T.~Quella, I.~Runkel and G.M.T.~Watts, \emph{{Reflection and Transmission for
  Conformal Defects}},
\doi{10.1088/1126-6708/2007/04/095}{{\emph{Journal of High
  Energy Physics} {2007} (2007) 095}},
[\href{https://arxiv.org/abs/hep-th/0611296}{{hep-th/0611296}}].


%3
\bibitem{Petkova:2000ip}
V.B.~Petkova and J.B.~Zuber, 
{\it Generalised twisted partition functions}, 
\doi{10.1016/S0370-2693(01)00276-3}{Phys.\ Lett.\ {B504} (2001)}
157--164 \arxiv{hep-th/0011021}{[hep-th/0011021]}
%%CITATION = HEP-TH 0011021;%%

%4
\bibitem{FFRS}
J.~Fr\"ohlich, J.~Fuchs, I.~Runkel and C.~Schweigert,
\emph{Duality and defects in rational conformal field theory},
\doi{10.1016/j.nuclphysb.2006.11.017}{Nucl.~Phys.~B763 (2007) 354-430}
[\href{https://arxiv.org/abs/hep-th/0607247}{{hep-th/0607247}}]

%5
\bibitem{Oshikawa:1996dj}
M.~Oshikawa and I.~Affleck, {\it Boundary
conformal field theory approach to the critical two-dimensional
Ising model with a defect line}, 
\doi{10.1016/S0550-3213(97)00219-8}{Nucl.\ Phys.\ B {495} (1997)}
533--582 \arxiv{cond-mat/9612187}{[cond-mat/9612187]}.
%%CITATION = NUPHA,B495,533;%%
 
%6
\bibitem{krw}
M.~Kormos, I.~Runkel and G.M.T.~Watts,
\emph{Defect flows in minimal models},
\doi{10.1088/1126-6708/2009/11/057}{{\emph{Journal of High Energy Physics, Volume 2009, JHEP11(2009)}}}
[\href{https://arxiv.org/abs/0907.1497}{{arXiv/0907.1497}}]

%7
\bibitem{mw}
I.~Makabe and G.M.T.~Watts
\emph{Defects in the Tri-critical Ising model},
\doi{10.1007/JHEP09(2017)013}{{\emph{J. High Energ. Phys. (2017) 2017:13}}}
[\href{https://arxiv.org/abs/1703.09148}{{arXiv/1703.09148}}]

%8
\bibitem{GY}
D.~Gang and S.~Yamaguchi, \emph{{Superconformal defects in the tricritical
  Ising model}},
  \doi{10.1088/1126-6708/2008/12/076}{\emph{Journal of High
  Energy Physics} {2008} (2008) 076},
  [\href{https://arxiv.org/abs/0809.0175}{{arXiv/0809.0175}}].

%\bibitem{Bachas:2001vj}
%C.~Bachas, J.~de Boer, R.~Dijkgraaf and
%H.~Ooguri, {\it Permeable conformal walls and holography}, JHEP {\bf
%0206} (2002) 027 \arxiv{hep-th/0111210}{[hep-th/0111210]}.

%9
\bibitem{G1}
D.~Gaiotto, \emph{{Domain Walls for Two-Dimensional Renormalization Group
  Flows}}, 
\doi{10.1007/JHEP12(2012)103}{\emph{Journal of
  High Energy Physics} {2012} (2012) 103},
  [\href{https://arxiv.org/abs/1201.0767}{{arXiv/1201.0767}}].

%\bibitem{Bachas:2007td}
%C.~Bachas and I.~Brunner, {\it Fusion of conformal
%interfaces}, JHEP {\bf 0802} (2008) 085 \arxiv{0712.0076}{[0712.0076 [hep-th]]}.


%\bibitem{Bachas:2004sy}
%C.~Bachas and M.R.~Gaberdiel, {\it Loop
%operators and the Kondo problem}, JHEP {\bf 0411} (2004) 065
%\arxiv{hep-th/0411067}{[hep-th/0411067]}.

%10
\bibitem{SCB}
I.~Brunner and C.~Schmidt-Colinet,
\emph{Reflection and transmission of conformal perturbation defects}
\doi{10.1088/1751-8113/49/19/195401}{\emph{J. Phys. A: Math. Theor. 49 (2016) 195401}}
[\href{https://arxiv.org/abs/1508.04350}{arXiv/1508.04350}]

%2b
\bibitem{BGLM}
M.~Bill\'o, V.~Gon\c calves, E.~Lauria and M.~Meineri,
\emph{Conformal Defects},
\href{https://doi.org/10.1007/JHEP04(2016)091}{J. High Energ. Phys. {(2016)} 2016: 91.}
[\href{https://arxiv.org/abs/1601.02883}{arxiv.org/1601.02883}]

%17
\bibitem{RCFTIV}
J.~Fuchs, I.~Runkel and C.~Schweigert,
{\it TFT construction of RCFT correlators IV:: Structure constants and correlation functions},
\doi{10.1016/j.nuclphysb.2005.03.018}{Nucl. Phys. B 715 (2005) 539--638}
[\arxiv{hep-th/0412290}{hep-th/0412290}]

%14b
\bibitem{IR10}
I.~Runkel,
\emph{Non-local conserved charges from defects in perturbed conformal
  field theory},
\doi{10.1088/1751-8113/43/36/365206}{J.~Phys.~A43:365206,2010}
[\arxiv{arXiv:1004.1909}{arXiv:1004.1909}]

%11
\bibitem{bhw}
Z.~Bajnok, L.~Hollo and G.M.T.~Watts,
\emph{Defect scaling Lee-Yang model from the perturbed DCFT point of view},
\doi{10.1016/j.nuclphysb.2014.06.019}{\emph{Nucl. Phys. B 886 (2014) 93--124}}
[\href{https://arxiv.org/abs/1307.4536}{{arXiv/1307.4536}}]

%12
\bibitem{CIZ}
A.~Cappelli, C.~Itzykson and J.-B.~Zuber,
{\it Modular invariant partition functions in two dimensions},
\doi{10.1016/0550-3213(87)90155-6}{Nucl. Phys. B 280 (1987) 445--465}.

%13
\bibitem{Graham}
K.~Graham and G.M.T.~Watts,
{\it Defect Lines and Boundary Flows},
\doi{10.1088/1126-6708/2004/04/019}{JHEP 0404 (2004) 019}
[\arxiv{hep-th/0306167}{hep-th/0306167}]

%14
\bibitem{runkelPhD}
I.~Runkel,
\emph{Boundary Problems in Conformal Field Theory},
PhD thesis, King's College London, 2000.


%15
\bibitem{Affleck:1991tk}
I.~Affleck and A.W.W.~Ludwig,
{\it Universal noninteger `ground state degeneracy' in 
critical quantum systems},
\href{10.1103/PhysRevLett.67.161}{Phys.\ Rev.\ Lett.\  {67} 161 (1991)}.
%%CITATION = PRLTA,67,161;%%

%16
\bibitem{Recknagel:2000ri}
A.~Recknagel, D.~Roggenkamp and V.~Schomerus,
{\it On relevant boundary perturbations of unitary minimal models},
\href{https://doi.org/10.1016/S0550-3213(00)00519-8}{Nucl.\ Phys.\ B {588} (2000) 552}
[ \arxiv{hep-th/0003110}{[hep-th/0003110]}]
%%CITATION = NUPHA,B588,552;%%


%18
\bibitem{DF}
Vl.S.~Dotsenko and V.A.~Fateev,
{\it Operator algebra of two-dimensional conformal theories with central charge $C \leq 1$},
\doi{10.1016/0370-2693(85)90366-1}{Phys.~Lett.~B4 (1985) 291--295}

%19
\bibitem{runkel}
I.~Runkel
\emph{Boundary structure constants for the A-series Virasoro minimal models},
\href{https://doi.org/10.1016/S0550-3213(99)00125-X}{{Nucl. Phys. B 549 (1999) 563--578}}
[\href{https://arxiv.org/abs/hep-th/9811178}{{hep-th/9811178}}]



\end{thebibliography}
%\begin{thebibliography}
%\renewcommand{\href}[2]{#2}
\newcommand\arxiv[2]      {\href{http://arXiv.org/abs/#1}{#2}}
\newcommand\doi[2]        {\href{http://dx.doi.org/#1}{#2}}
\newcommand\httpurl[2]    {\href{http://#1}{#2}}

\begingroup\raggedright\endgroup

\end{document}